\documentclass[aip,graphicx,reprint]{revtex4-1}

\usepackage{graphicx}
\usepackage{dcolumn}
\usepackage{bm}
\usepackage{amsmath}
\usepackage{amssymb}
\usepackage{siunitx}
\usepackage[T1]{fontenc}
\usepackage{mathptmx}
\usepackage{etoolbox}
\usepackage{99_IMS_constants}

\DeclareSIUnit{\belmilliwatt}{Bm}
\DeclareSIUnit{\dBm}{\deci\belmilliwatt}
\DeclareSIUnit{\pH}{\pico\henry}

\makeatletter
\def\@email#1#2{%
 \endgroup
 \patchcmd{\titleblock@produce}
  {\frontmatter@RRAPformat}
  {\frontmatter@RRAPformat{\produce@RRAP{*#1\href{mailto:#2}{#2}}}\frontmatter@RRAPformat}
  {}{}
}
\makeatother
\begin{document}

\title{Simulation framework for microwave SQUID multiplexer optimization}

\author{C. Schuster}%
 \email{constantin.schuster@kit.edu}
\affiliation{Institute of Micro- and Nanoelectronic Systems, Karlsruhe Institute of Technology, Hertzstrasse 16, Building 06.41, D-76187 Karlsruhe, Germany}

\author{M. Wegner}%
\affiliation{Institute of Micro- and Nanoelectronic Systems, Karlsruhe Institute of Technology, Hertzstrasse 16, Building 06.41, D-76187 Karlsruhe, Germany}
\affiliation{Institute for Data Processing and Electronics, Karlsruhe Institute of Technology, Hermann-von-Helmholtz-Platz 1, Building 242, D-76344 Eggenstein-Leopoldshafen}

\author{S. Kempf}%
\affiliation{Institute of Micro- and Nanoelectronic Systems, Karlsruhe Institute of Technology, Hertzstrasse 16, Building 06.41, D-76187 Karlsruhe, Germany}
\affiliation{Institute for Data Processing and Electronics, Karlsruhe Institute of Technology, Hermann-von-Helmholtz-Platz 1, Building 242, D-76344 Eggenstein-Leopoldshafen}

\date{\today}

\begin{abstract}
So far, performance prediction and optimization of microwave SQUID multiplexers has largely been based on simple approximate analytical models and experimental results. This is caused by the complexity of the underlying physics and the intricacy of operation and readout parameters. As a simplified description can never account for all potential effects occurring in a real device, we have developed a software framework to simulate the characteristics and performance of a microwave SQUID multiplexer. Our simulation framework is a powerful tool to guide understanding and optimization of microwave SQUID multiplexers and other related devices. It includes common readout schemes such as open-loop or flux ramp modulated readout as well as the nonlinear behavior of Josephson tunnel junctions. Moreover, it accounts for the non-zero response time of superconducting microwave resonators with high loaded quality factors as well as the most significant noise contributions such as amplifier noise, resonator noise as well as SQUID noise. This ultimately leads to a prediction of device performance that is significantly better as compared simple analytical methods. Using the simulation framework, we discuss first steps towards a full microwave SQUID multiplexer optimization and highlight some other applications which our simulation framework can be used for.
\end{abstract}

\maketitle

\section{Introduction}
\label{sec:intro}

Cryogenic detectors such as superconducting transition-edge sensors (TESs) \cite{Irwin2005,Ullom2015},  magnetic microcalorimeters (MMCs) \cite{Fleischmann2005,Kempf2018} or magnetic penetration depth thermometers (MPTs) \cite{Nagler2012, Bandler2012} have impressively proven to be among the most sensitive devices for measuring incident power or energy. For this reason, they represent the current state of the art for bolometric or calorimetric applications. Various experiments strongly benefit from or even rely on the exceptional and outstanding properties of these detectors. Using an ultra-sensitive thermometer, based on superconducting (TES, MPT) or paramagnetic (MMC) materials, as well as an appropriate low-impedance readout circuit, they convert the actual input signal into a change of electrical current or magnetic flux that is continuously measured with utmost sensitivity by means of a wideband superconducting quantum interference device (SQUID) \cite{Fagaly2006}.

The maturity of fabrication technology allows 'easily' building detector arrays of virtually any size. Out of these, small-scale detector arrays with up to a few tens of detectors can be readily read out with individual single-stage or two-stage dc-SQUIDs as they are used for single-channel readout. In contrast, medium-scale and particularly large-scale detector arrays necessarily demand the usage of cryogenic SQUID based multiplexing techniques to address the challenging requirements related to overall cost, system complexity and the interplay between readout induced power dissipation and cooling power of the cryostat. 

\begin{figure}
  \includegraphics[width=0.8\columnwidth]{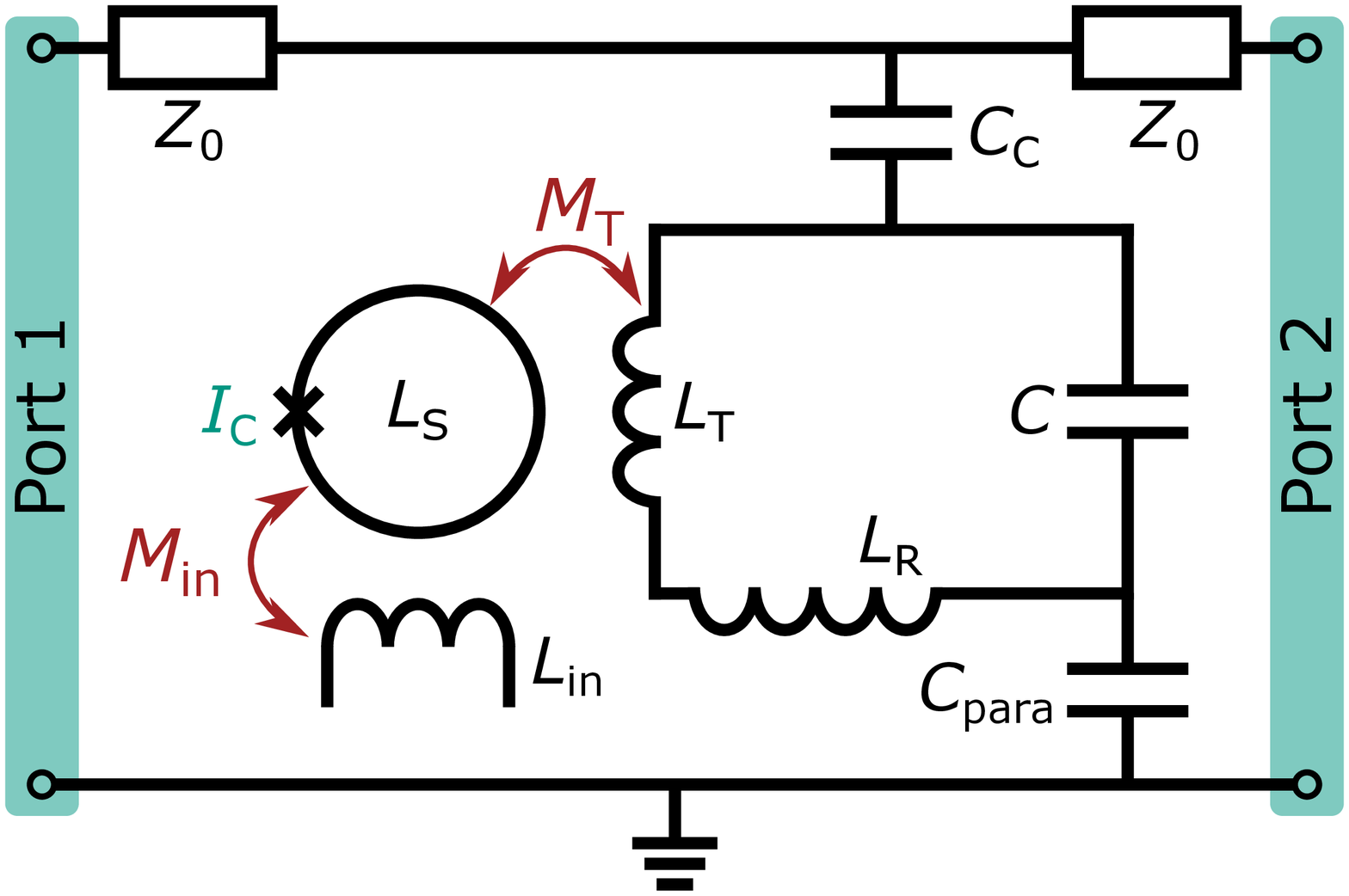}
  \caption{Simplified schematic circuit diagram of a single readout channel of a lumped element based microwave SQUID multiplexer.}
  \label{fig:lemwr_schem}
\end{figure}

Existing SQUID multiplexers rely on time-division \cite{Doriese2016}, frequency-division using MHz \cite{Hartog2014, Richter2021} or GHz carriers \cite{Mates2008,Hirayama2013, Kempf2017}, code-division \cite{Morgan2016} or hybrid \cite{Reintsema2008, Irwin2018, Yu2020, HyMUX2022} multiplexing schemes. Out of these, microwave SQUID multiplexing \cite{Mates2008,Hirayama2013, Kempf2017} appears to be best suited for the readout of large and ultra-large scale detector arrays as the bandwidth per readout channel does not necessarily have to be restricted and readout noise is to first order independent of the number of readout channels.
A microwave SQUID multiplexer ({\uMUX}) employs transmission line or lumped element based superconducting microwave resonators as frequency encoding elements. Each resonator is capacitively coupled to a transmission line, common to all readout channels of the multiplexer, and inductively coupled to a non-hysteretic current-sensing rf-SQUID being connected to the associated cryogenic detector. Due to its parametric self-inductance, the SQUID transduces the detector signal into a change of amplitude and phase of a microwave signal continuously probing the resonance frequency of the resonator. Figure~\ref{fig:lemwr_schem} shows a simplified equivalent circuit diagram of a single {\uMUX} readout channel based on a lumped element resonator. The resonator is formed by the parallel circuit consisting of the capacitance $C$ and the inductance $L = L_\mathrm{R} + \LT$. It is coupled to a transmission line with impedance $Z_0$ via the capacitance $\CC$ and coupled to ground by a parasitic capacitance $\Cpar$. The effects of this parasitic capacitance can be described by an effective value $\CCeff = (\CC^{-1}+\Cpar^{-1})^{-1}$ for the coupling capacitance. 
The load inductance $\LT$ inductively couples the resonator to the SQUID with mutual inductance $\MT = \kT \sqrt{\LT \LS}$. Here, $k_\mathrm{T}$ denotes the geometrical coupling factor. The SQUID comprises a closed superconducting loop with inductance $\LS$ that is interrupted by a single unshunted Josephson tunnel junction with critical current $\Ic$. To guarantee non-hysteretic, i.e. dispersive, operation, the SQUID screening parameter is $\betaL = 2\pi \LS \Ic / \PhiO < 1$. A current $I_\mathrm{in}$ running through the input coil with inductance $\Lin$, as caused by a detector signal, induces a magnetic flux signal $\Phi_\mathrm{in} = M_\mathrm{in} I_\mathrm{in}$
threading the SQUID loop. In this arrangement, the resonance frequency is altered as the flux through the SQUID loop changes.
The former can easily be read out by applying a fixed microwave probe tone and measuring amplitude and/or phase of the transmitted signal.

The periodicity of the magnetic flux dependent resonance frequency (the period is given by the magnetic flux quantum $\PhiO$) necessitates a method for linearizing the {\uMUX} output signal. The most common method is flux ramp modulation (FRM)\cite{Mates2012}. Here, a sawtooth-shaped current signal is injected into a modulation coil with inductance $\Lmod$, which is connected in series with the corresponding coils of the other channels. The modulation coil is inductively coupled to the SQUID via the mutual inductance $\Mmod = k_\mathrm{mod}\sqrt{\Lmod \LS}$. For each ramp cycle, a linearly increasing flux bias is induced. Amplitude $I_\mathrm{mod}^\mathrm{max}$ and repetition rate $\framp$ of the modulation signal are chosen such that an integer number of flux quanta are induced in the SQUID loop and that the detector signal is quasi-static within a cycle of the flux ramp. In this case, the detector signal manifests as a phase offset in the periodic SQUID response that is proportional to the input signal \cite{Mates2012}.

Due to the non-linearity of the Josephson equations describing the underlying physics of Josephson tunnel junctions as well as the associated dependence of the SQUID response on probe tone power \cite{Wegner2022}, {\uMUX} characteristics are intrinsically non-linear. Additional non-linear effects arise from the non-zero resonator response time and interdependencies of {\uMUX} parameters as, for example, readout power and resonance frequency. In combination with noise emerging from passive and active components of the microwave setup as well as the complexity of the FRM readout, this leads to an intricate physical behavior significantly complicating or even preventing the application of analytical methods for {\uMUX} description and optimization. However, as the optimization of design and readout parameters is crucial for next-generation detector systems, we have developed a simulation framework to explore and optimize {\uMUX} behavior by means of numerical simulations.

In this paper, we describe the structure of our simulation framework. This includes a short review and discussion of the used physics models and numerical algorithms as well as a summary of the input parameters and settings that need to be specified for performing a simulation run. We then show that our simulation results are in very good agreement with expectations based on information theory as well as experimental data. We explicitly show that our simulations describe acquired data much better than existing analytical models, which are unable to account for all interdependencies and non-linear effects. Finally, we outline possible areas of applications of our simulation framework. This includes an analysis of the remaining nonlinearity between the input and output signal despite the use of flux ramp modulation as well as a first step towards full {\uMUX} optimization. The latter is, however, not within the scope of this paper and will be presented and discussed in a future publication.

\section{Description of the simulation framework}\label{sec:uMUXing_concept}

The physics of a microwave SQUID multiplexer is governed by several implicit equations that can hardly be tackled by analytical means. For this reason, we apply numerical methods to assess {\uMUX} characteristics and performance for a predefined set of device parameters. More precisely, we generate/calculate a time-discrete transmission time trace $\Stok = \Sto(t_k)$ with $k = 0,...,N-1$ and $N \in \mathbb{N}$ at equidistant points $t_k$ in time. This time trace $\Stok$ represents a discrete version of the time-dependent, complex-valued transmission parameter $\Sto(t)$ of a single {\uMUX} channel as sampled in a real setup using a data acquisition system running with sampling rate $\fs = 1/(t_{k} - t_{k-1})$. We then treat this artificial time trace in the same way as experimental data to yield, for example, a magnetic flux noise spectrum.

\begin{figure*}
\centering
  \includegraphics[width=0.7\textwidth]{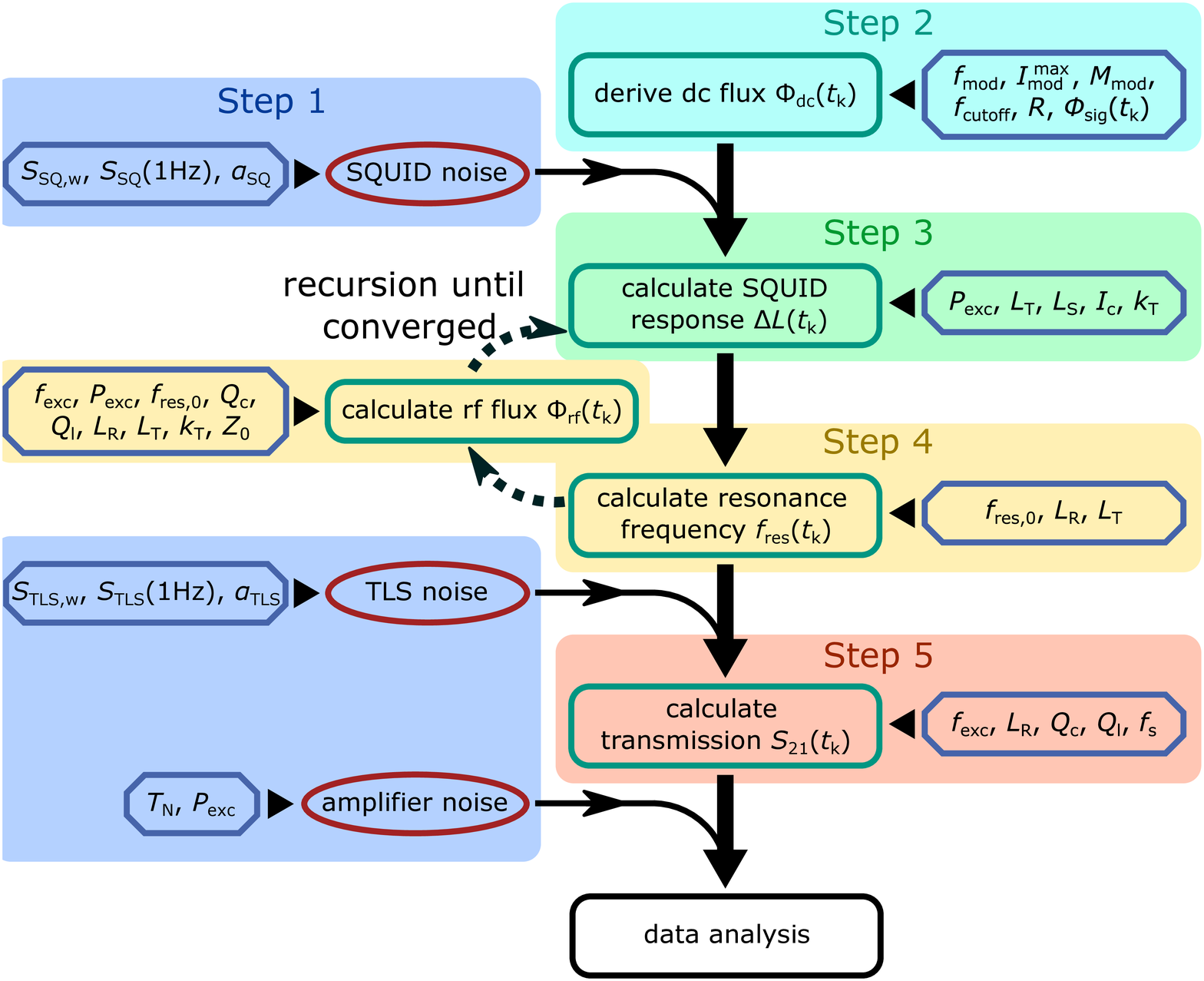}
  \caption{Flowchart outlining the structure of our  simulation framework. Blue octagons represent user specified input parameters, red ovals depict noise generation and green boxes general calculations. The individual steps as well as the meaning of the different symbols and variables are explained in the main text.}
  \label{fig:numsim_schem}
\end{figure*}

Figure~\ref{fig:numsim_schem} depicts a flowchart outlining the structure of our simulation framework to perform a single simulation run yielding the transmission time trace $\Sto(t_k)$ for a given set of device and readout parameters. This time trace is then analyzed using a modified Welch's method (for details see section~\ref{sec:MoCaSim}) to determine the magnetic flux noise spectral density. In the following, we give a short overview of the basic workflow of such a single simulation run in chronological order. In section~\ref{sec:MoCaSim}, we then comprehensively discuss the individual steps including all specifics and underlying equations.

\subsection{Step 1: Generation of noise traces}
\label{ssec:gennoise}
The first step is the generation of quasi-random noise time traces. We include three noise sources, i.e. amplifier noise added along the entire output signal path, two-level system (TLS) noise of the readout resonators affecting the resonance frequency as well as magnetic flux noise of the SQUID (SQ). We assume the amplifier noise to be white, i.e. frequency-independent. Its magnitude is calculated according to the predefined effective noise temperature $\TN$ of the readout system as well as the readout power $\Pexc$. For both, TLS noise and SQUID magnetic flux noise, we assume the noise to be composed of a frequency-independent white and a frequency-dependent $1/f^\alpha$-like contribution. Either noise trace is hence generated according to three input parameters $S_{i,\mathrm{w}}$, $S_{i}(1\,\mathrm{Hz})$ and $\alpha_i$ with $i \in \{\mathrm{TLS, SQ}\}$ determining the resulting noise spectral density $S_i = S_{i,\mathrm{w}} + S_i(1\,\mathrm{Hz})/f^\alpha$. As such, $S_{i,\mathrm{w}}$ represents the amplitude of the white noise contribution and $S_{i}(1\,\mathrm{Hz})$ the amplitude at a frequency of $f = 1\,\mathrm{Hz}$ and the exponent $\alpha$ of the $1/f^\alpha$-like contribution. It is worth mentioning that changing the shape of the noise spectra of either noise contribution requires only minor modifications of the source code, thus the shape of the noise spectrum can be easily adapted to actual experimental data.

\subsection{Step 2: Calculation of the quasi-static magnetic flux threading the the SQUID loop}
\label{ssec:dcflux}
The magnetic flux threading the SQUID loop is composed of three contributions, i.e. the actual input signal $\Phi_\mathrm{in}$, the sawtooth-shaped flux ramp $\Phi_\mathrm{mod}$ as well as the magnetic flux $\Phi_\mathrm{rf}$ induced by the microwave signal within the readout resonator. Out of these, the input signal and the flux ramp appear to be quasi-static as compared to the flux induced by the microwave signal. For this reason, we denote the first two contributions as 'dc-flux', though it is slowly (with respect to the microwave signal) changing over time. 

The magnetic flux $\Phidc = \Phiin + \Phimod$ is composed of the preset noise-free input signal, the noise-free flux ramp signal as well as the flux noise time trace derived in the previous step (see section~\ref{ssec:gennoise}). Both, open-loop and FRM readout, can be modeled using the simulation framework, depending on the chosen input parameters. For open-loop readout, the modulation signal takes a predefined constant value representing a static magnetic flux bias. For flux ramp modulation, the flux signal is time-dependent and takes the shape of a sawtooth signal with ramp reset rate $\framp$ and amplitude $\Phimodmax = \Mmod\Imodmax$. Here, $\Imodmax$ denotes the amplitude of the current running through the modulation coil and $\Mmod$ the mutual inductance between SQUID and modulation coil. Optionally, a Butterworth lowpass filter with predefined filter order $R$ and cutoff frequency $f_\mathrm{cutoff}$ can be applied to the flux ramp signal to mimic a real system with finite bandwidth.

\subsection{Step 3: Derivation of the effective inductance shift}
Our simulation framework provides three methods to calculate the time-dependent effective change $\dLTk$ of the inductance of the readout resonator as caused by the SQUID. More precisely, it allows to choose between two analytical equations for the limiting cases of either very weak, i.e. negligible probe tone power $\Pexc$, or very small, i.e. vanishing screening currents in the SQUID loop. The first scenario can be applied for arbitrary values of the screening parameter $\betaL < 1$ as long as $\Pexc \rightarrow 0$, the second method is valid for any value of the probe tone power $\Pexc$ as long as $\betaL \ll 1$. However, in most cases and in particular in situations relevant for real applications neither of both situations applies. For this reason, our simulation framework provides a third method that takes into account both non-zero values of the probe tone power $\Pexc$ and non-zero values of the screening parameter up to $\betaL \approx 0.6$. The latter method is based on our most recent microwave SQUID multiplexer model \cite{Wegner2022}. In this step, the SQUID inductance $\LS$, the critical current $\Ic$, the geometric coupling parameter $\kT$ between resonator and SQUID as well as the mutual inductance $\Mmod$ between modulation coil and SQUID are input parameters for the simulation. In addition, the dc flux $\Phidc$ within in the SQUID loop (see section~\ref{ssec:dcflux}) derived in the previous step as well as the magnetic flux contribution $\Phirf$ induced by the current flowing within the resonator enter. It is worth mentioning that for the second and third method, the implicit nature of the underlying equations (see sections~\ref{subsec:MoSimVanCur} and~\ref{sec:numsim_general}) requires to use an iterative numerical approach to calculate an accurate prediction of the time-dependent magnetic flux contribution as caused by the microwave currents within the resonator.

\subsection{Step 4: Calculation of the actual resonance frequency}
The time-dependent resonance frequency $\fresk$ is calculated using the effective inductance shift $\dLT$, the predefined resonator parameters as well as the generated noise trace due to TLS noise (see section~\ref{sec:MoCaSim}). Here, the predefined resonator parameters are the unloaded resonance frequency $\fresO$, the resonator inductance $\LR$, the coupling inductance $\LT$, the coupling quality factor $\Qc$, the loaded quality factor $\Ql$ and the impedance $Z_0$ of the transmission line. 
This step may have to be performed iteratively for non-zero values of the probe tone power $\Pexc$ (see sections~\ref{subsec:MoSimVanCur} and~\ref{sec:numsim_general})as the effective inductance shift $\dLT$ depends on the microwave power stored in the resonator. However, the latter depends on the resonance frequency $\fres$ which in turn depends on the effective inductance shift $\dLT$.
Once the resonance frequency $\fres$ is calculated, the effective resonance frequency noise caused by TLS is added to yield the time trace of the resonance frequency $\fresk$.

\subsection{Step 5: Derivation of the transmission coefficient}
The final step of a single simulation run is the calculation of the transmission $\Stok$. Since the modulation of the resonance frequency, especially for FRM readout, can be rather fast, a steady-state approximation for the resonator response is no longer applicable. For this reason, we consider the non-equilibrium dynamics of the resonator response using a first order approximation. After deriving the complex transmission coefficient $\Stok$, amplifier noise is added. This yields the final simulation output that is afterwards treated in the same way as experimental data.
\section{Detailed description of the simulation framework} \label{sec:MoCaSim}

The first step of a simulation run is the generation of the noise time traces. The method used to generate noise is identical for all three sources, solely the power spectral density differs. The goal is to synthesize a random discrete time trace $x_k$ of noise at discrete points $t_k = k/\fs$ in time, with $k = 0,...,N-1$ and $N \in \mathbb{N}$, based on a given noise spectral density of $\hat{S}_\mathrm{x}(f)$. Here, $\fs$ is the rate at which the signal is sampled. For this, noise coefficients $\hat{a}_j$ in frequency space are generated with amplitudes
\begin{eqnarray}\label{eq:noise01}
\left| \hat{a}_j \right| = 
\begin{cases}
\sqrt{\hat{S}_\mathrm{x} (\fs \frac{j}{N})}\ & \text{for}\ 
j = -\frac{N}{2},-\frac{N}{2}+1,...,\frac{N}{2}-1\\
0\ & \text{for}\ j = 0\\
\end{cases}
\end{eqnarray}
as well as random phases $\theta_j$ following a uniform distribution
\begin{equation}\label{eq:noise02}
\hat{a}_j = \left| \hat{a}_j \right| \rme^{\rmi \theta_j},\quad
\theta_j \in \left[0,2\pi\right).
\end{equation}
The noise coefficient $\hat{a}_0$ at zero frequency must vanish to ensure zero-mean noise, regardless of the targeted noise spectral density.
Using an inverse fast Fourier transform yields a complex-valued discrete noise time trace:
\begin{equation}\label{eq:noise03}
x_k = \sqrt{\frac{f_s}{2}} \sum_{j=-N/2}^{N/2-1} \rme^{2\pi \rmi \frac{jk}{N}} \hat{a}_j,\quad k = 0,...,N-1.
\end{equation}
In case that real-valued noise is needed, the sum of the real and imaginary contributions of each $x_k$ is used. The spectral density of the noise remains the same. In the simulations, the transmission noise caused by the amplifier is complex-valued, whereas the flux noise in the SQUID and the resonance frequency noise are real-valued.

The noise time traces of the various sources are subsequently included into the generation of the transmission data, along with a set of device- and readout parameters as well as a signal time trace defining the input flux into the SQUID loop. The external flux contribution $\phidck = \phisigk + \phimodk + \dphik$ (from here on we use normalized magnetic flux values, i.e. $\varphi \equiv 2\pi\Phi/\Phi_0$) is the sum of the flux signal time trace $\phisigk$, the modulation flux $\phimodk$ and the magnetic flux noise $\dphik$. Each of these contributions is assumed to be quasi-static with respect to the resonance frequency $\fres$.

In a microwave SQUID multiplexer operated with flux ramp modulation, a sawtooth-shaped modulation current $\Imod$ is applied to the modulation coil. The mutual inductance between SQUID loop and modulation coil is $\Mmod$, leading to a modulation flux $\phimod = 2\pi \Mmod \Imod / \PhiO$. In the simulation framework, a modulation current time trace $\Imodk$ is generated using a predefined ramp repetition rate $\framp$ and ramp amplitude $\Imodmax$. In software, a sawtooth shape with infinitely steep resets and perfectly linear ramp segments can be generated. However, to mimic real electronics, we include to possibilty to apply a Butterworth lowpass filter may be applied to the time trace $\Imodk$ to emulate the finite bandwidth of real signal generators and transmission lines. 
If the simulation is run with open-loop readout, the modulation flux is assumed to be constant, i.e. $\phimodk = \phibias = \mathrm{const.}$. The bias flux is then typically chosen such that the transfer coefficient $\Kphi(\Phi) =  \left( \partial \left| \Sto(\Phi) \right| / \partial \Phi \right)$ is maximised: $\Kphi (\phibias) = K_\mathrm{\Phi}^\mathrm{max}$. 

The method to derive the time trace $\fresk$ of the resonance frequency depends on the actual device parameters. This results in different expressions for the inductance shift $\dLT$ with varying numerical complexity. We hence choose the actual method on the basis of the predefined device parameters.

\subsection{Vanishing probe tone power $\Phirf \to 0$}
\label{subsec:Pexc_to_zero}

For vanishing probe tone power, i.e. $\Phirf \to 0$,  an analytic solution for the inductance shift $\dLT$ exists. It is given by the expression\cite{Wegner2022}
\begin{equation} \label{eq:dLT_phirf0}
\dLT = \frac{\MT^2}{\LS} \frac{\betaL \cose{\phitot}}{1 + \betaL \cose{\phitot}}.
\end{equation}
To calculate the inductance shift $\dLT$, the total magnetic flux $\phitot$ threading the SQUID loop must be determined. Due to screening currents within in the SQUID loop $\phitot$, the latter is given by the expression \cite{Wegner2022}
\begin{equation}\label{eq:phitot_of_phidc}
    \phitot = \phidc - \betaL \sine{\phitot}.
\end{equation}
Despite being an implicit equation, this relation is unique for $\betaL < 1$ and can be inverted to yield the explicit expression
\begin{equation}\label{eq:phidc_of_phitot}
    \phidc = \phitot + \betaL \sine{\phitot}
\end{equation}
which is evaluated at $1000$ linearly spaced data points $\phitotj \in [0,2\pi)$ yielding an equal number of points $\phidcj$. Since the relation is unique, a cubic spline interpolation to the dataset can be performed, yielding an interpolation function $f(\phidc)$ such that $f(\phidcj) = \phitotj$. Moreover, as $\dLT (\phitot)$ is $2\pi$-periodic, the restriction to nodes $\phitotj$ on the interval $[0,2\pi)$ is sufficient. Using the interpolation function $f(\phidc)$, we calculate the $\phitotk$ for each value of $\phidck$ for the given time trace which is used in the subsequent evaluation of equation \ref{eq:dLT_phirf0} to obtain an inductance shift time trace $\dLTk$. Once the inductance shift has been calculated, the resulting resonance frequency time trace $\fresk$ is derived using the expression
\begin{equation} \label{eq:fres_reults}
\fresk = \fresO \left( 1 - \frac{\dLTk}{\LR + \LT}\right) ^{-\frac{1}{2}}.
\end{equation}

\subsection{Vanishing screening currents $\betaL \to 0$}\label{subsec:MoSimVanCur}
For vanishing screening currents within the SQUID loop, the analytic solution
\begin{equation}\label{eq:dLt_betaL0}
    \dLT = \frac{\MT^2 \betaL}{\LS} \frac{2 J_1(\phirf)}{\phirf}\cose{\phidc}
\end{equation}
for the inductance shift $\dLT$ exists\cite{Wegner2022}. Here, the flux amplitude $\phirf$ enters  that is caused by the microwave current running within the inductor $\LT$. It thus depends on the energy stored within the resonator that  in turn depends on the relative position between the resonance frequency $\fres$ and the frequency of the probe tone $\fexc$. Since the resonance frequency depends on $\dLT$, the implicit equation~\ref{eq:dLt_betaL0} can't be solved directly. The radio frequency flux amplitude is given by $\phirf = 2 \pi \MT \IT / \PhiO$. Here, $\MT = \kT \sqrt{\LT \LS}$ denotes the mutual inductance between SQUID loop and the load inductor and $\IT$ is the amplitude of the microwave current running in the inductor $\LT$. The latter is calculated using the analytical expression\cite{Ahrens_Diss}
\begin{widetext}
\begin{equation} \label{eq:IT_offreso}
\IT(\fexc) =  \sqrt{2 \Pexc \Z0}
\frac{2 \pi \fexc \sqrt{\frac{2}{\Z0 (2 \pi \fres)^3 L \Qc}}}
{\left( 2i - 2 \pi \fexc \sqrt{\frac{2}{\Z0 (2 \pi \fres)^3 L \Qc}} \Z0 \right)
\left( \frac{\fexc^2}{\fres^2} - 1 \right) + \frac{\fexc^3}{\fres^3} \frac{2}{\Qc}}.
\end{equation}
\end{widetext}
To derive the resonance frequency time trace $\fresk$, we evaluate equation~\ref{eq:dLt_betaL0} assuming a vanishing radio frequency flux $\phirfk^0 = 0$. For this, we derive a first guess $\fresk^0$ using equation~\ref{eq:fres_reults}. Using this guess, we calculate a more accurate guess $\phirfk^1$ for the amplitude of the radio frequency flux using equation~\ref{eq:IT_offreso}. These steps are repeated until subsequent results for the resonance frequency have a sufficiently small deviation:
\begin{eqnarray}\label{eq:itersetp_fresk}
\phirfk^0 &= 0 \quad \forall\ k,\\
\hfresk^m &= \widetilde{\fres}(\phidck,\phirfk^{m}),\\
\phirfk^{m+1} &= \widetilde{\phirf}(\hfresk^m,\fexck).
\end{eqnarray}
up until 
\begin{equation}\label{eq:itersetp_fresk_breakcond}
\sum_{k=0}^{N}\frac{\hfresk^M - \hfresk^{M-1}}{\hfresk^M} \leq \epsilon_\mathrm{f}
\end{equation}
at some $M \in \mathbb{N}$ for a given maximum tolerable difference $\epsilon_\mathrm{f}$. The result $\fresk \equiv \fresk^M$ is the resonance frequency time trace used for the remaining part of the simulation run. In this description, $\widetilde{\fres}(\phidc,\phirf)$ refers to equations \ref{eq:fres_reults} and \ref{eq:dLt_betaL0}, and $\widetilde{\phirf}(\fres,\fexc)$ follows from equation \ref{eq:IT_offreso}. By design, this method only works if $\hfresk^m$ is converging. This has been the case for all reasonable choices of simulation parameters we have tested so far. 

\subsection{General case}\label{sec:numsim_general}
In general, both the screening parameter $\betaL$ and the probe tone power $\Pexc$ take non-zero values. For describing the underlying physics, our most recent multiplexer model yields the expression \cite{Wegner2022}
\begin{equation}\label{eq:dLt_general}
    \dLT = \frac{\MT^2 \betaL}{\LS} \frac{2}{\phirf} \sum_{i,j}a_{i,j}\betaL^{b_{i,j}} J_1(c_{i,j}\phirf)\cose{c_{i,j}\phidc}
\end{equation}
which is valid for $\betaL \leq 0.6$. Here, $a_{i,j}$, $b_{i,j}$ and $c_{i,j}$ are coefficients that are listed in \cite{Wegner2022}.
To derive the inductance shift $\dLT$ in the general case, we apply the same recursive method as described in section~\ref{subsec:MoSimVanCur}, with the difference that equation~\ref{eq:dLt_general} is used instead of equation~\ref{eq:dLt_betaL0}.

Once the resonance frequency time trace $\hfresk$ has been derived, an effective resonance frequency noise $\dfresk$ is added, representing the noise contribution of two-level systems within the resonator. The noisy resonance frequency trace $\fresk = \hfresk + \dfresk$ is then used to calculate the transmission time trace $\hStok$.

Assuming a sufficiently slow modulation of the resonance frequency, the transmission of a resonator can be approximated by the steady-state expression\cite{Zmuidzinas2012}
\begin{equation} \label{eq:S21_steadystate}
\Sto^\mathrm{SS}(t) \approx \frac{\frac{\Ql}{\Qi} + 2 \rmi \Ql \delta f(t)}{1 + 2 \rmi \Ql \delta f(t)}
\end{equation}
with the relative frequency difference $\delta f(t) = (\fexc - \fres(t))/\fres(t)$. However, in practice, the resonance frequency $\fres(t)$ changes rather fast such that it can not be approximated as quasi-static. For this reason, a dynamic resonator description has to be used which we approximated to first order by the expression
\begin{eqnarray} \label{eq:S21_iter001}
\Sto(t_0 + \Delta t) &\approx& \Sto^\mathrm{SS}(t_0 + \Delta t) \\
&+& \left[ \Sto(t_0) - \Sto^\mathrm{SS}(t_0 + \Delta t) \right] \rme^{-\pi \left[ \dfBW - 2 \rmi \left( \fres - \fexc \right)\right] \Delta t} \nonumber
\end{eqnarray}
as shown in appendix~\ref{sec:dynresmodel}. In case that the initial value of $\Sto(t_0)$ at a time $t_0$ is known, the transmission parameter $\Sto(t_0 + \Delta t)$ at time $t+\Delta t$ can be derived. Applying this method over and over again allows generating a time trace of arbitrary length. It is worth mentioning that this approximation is only valid assuming $\Sto^\mathrm{SS}$ to be quasi-static on the time scale $\Delta t$.
In the simulation, we use the steady state value $\hStoO = \Sto^\mathrm{SS}(t_0)$ for the initial time $t_0$ as starting value. The time interval $\Delta t = t_{k+1}-t_{k} = 1/\fs$ is given by the sampling rate $\fs$:
\begin{eqnarray} \label{eq:S21_iter002}
\hStoO &=& \hat{S}_{\mathrm{21},0}^\mathrm{SS}, \\
\hStokpo &=& \hat{S}_{\mathrm{21},k+1}^\mathrm{SS} \\ 
&+& \left( \hStok - \hStok^\mathrm{SS} \right) \rme^{-\pi \left[ \dfBW -  2 \rmi \left( \fresk - \fexck \right) \right] \Delta t}\nonumber.
\end{eqnarray}
Finally, transmission noise $\dStok$ is added, representing amplifier noise caused by the HEMT amplifier, yielding the final simulation output
\begin{equation} \label{eq:Stok}
\Stok = \hStok + \dStok.
\end{equation}
Here, the relation between the transmission noise spectral density $\hat{S}_\mathrm{S21}$ and the system noise temperature $\TN$ is given by:
\begin{equation} \label{eq:TNHEMT}
\sqrt{\hat{S}_\mathrm{S21}} = 2 \sqrt{\frac{2 \kB \TN}{\Pexc}}.
\end{equation}
This transmission time trace $\Stok$ resembles a measurement on a {\uMUX} device with the given parameters and can hence be treated in the same way as experimental data for subsequent analysis. For this reason,  demodulation of the transmission time trace yields the output signal flux $\phioutj$
\begin{equation} \label{eq:phisig_FRM}
\phioutj = \mathrm{arctan} \left[ \frac{\sum_{k = jW}^{(j+1)W - 1} \sine{2 \pi j \fmod/\fres } \left| \Stok \right|}
{\sum_{k = jW}^{(j+1)W - 1} \cose{2 \pi j \fmod / \fres} \left| \Stok \right|} \right],
\end{equation}
in case that flux ramp modulation is used. Here, $\fmod = \framp \Mmod \Imodmax / \PhiO$ denotes the modulation frequency $W = \fs/\framp$ is the number of data points in between two resets of the modulation ramp. Obviously, the resulting signal time trace $\phioutj$ has a factor of $W$ fewer points than the transmission time trace $\Stok$. For open-loop readout, the signal flux time trace $\phioutk$ can be calculated from the transmission time trace $\Stok$ using the transfer coefficient $K_\mathrm{\Phi} (\phibias)$:
\begin{equation} \label{eq:phisig_openloop}
\phioutk = \frac{\Stok}{K_\mathrm{\Phi} (\phibias)}.
\end{equation}
Here, the transfer coefficient $K_\mathrm{\Phi} (\phibias)$ is determined during the simulation by numerically calculating the transmission-to-flux characteristic $\Sto(\varphi_l)$ for 1024 linearly spaced data points of $\varphi_l \in [0,2\pi) $, and then subsequently calculating the numerical derivative at the specified bias flux value $\phibias$. The signal flux time trace has the same number of data points as the transmission time trace $\Stok$.

For noise analysis, e.g. to calculate the noise spectral density, a modified Welch's method \cite{Welch1967} is applied to the output signal. This method is based on the calculation of a number of $Q$ individual periodograms $P^q(f)$, each of which covers a subset of data points of the output signal time trace $\phioutk$. The length of these subsets $L$ must be smaller than the total number of data points in the output signal time trace $\phioutk$, and subsequent subsets overlap with $L-D$ datapoints. All $Q$ datasets combined cover the entirety of $\phioutk$. Each periodogram is then given by 
\begin{equation} \label{eq:Welch_1}
P^q(f) = \frac{2}{\fs \sum_{n=0}^{L-1} w_n^2} \left| \sum_{n=0}^{L-1} w_n \phi_{\mathrm{out},qD + n} \rme^{- 2 \pi \rmi f n /fs}\right|^2,
\end{equation}
with the weights $w_n$ of a window function. For the data presented in this paper, a Blackman-Harris window was used. The estimator $\Sphi(f)$ of the noise spectral density of the output signal time trace $\phioutk$ is then given by the average of all periodograms:
\begin{equation} \label{eq:Welch_2}
\Sphi(f) = \frac{1}{Q} \sum_{q=0}^{Q-1} P^q(f).
\end{equation}
If the length $L$ of the subsets is chosen large, the estimator $\Sphi(f)$ contains information even down to low frequencies $f$. However, the number $Q$ of individual sets is rather small, and only few individual periodograms can be averaged, leading to a low fidelity of the estimator. A choice of short window lengths $L$ results in many subsets and thus a high fidelity of the estimator, but the estimator can not resolve low frequencies. In this paper, we hence repeat this process for multiple different window lengths $L_i$. The combination of different subset lengths allows both a high estimator fidelity at large frequencies $f$ as well as information about low frequencies, albeit at a lower fidelity.

\section{Default simulation parameters}

Our simulation framework allows simulating the characteristics and performance of microwave SQUID multiplexers with virtually arbitrary input parameters. We presently only recommend that the input parameters should be taken from the parameter range for which our multiplexer model\cite{Wegner2022} has been approved, i.e. $\betaL \leq 0.6$, $Q_\mathrm{l} > 1000$ and, $4\,\mathrm{GHz} \leq \fresO \leq 8\,\mathrm{GHz}$. We presently investigate whether our model is still valid for higher resonance frequencies and work on model expansions to adequately describe adequately the multiplexer behavior for screening parameters $\betaL \rightarrow 1$.

\begin{figure}
  \includegraphics[width=0.8\columnwidth]{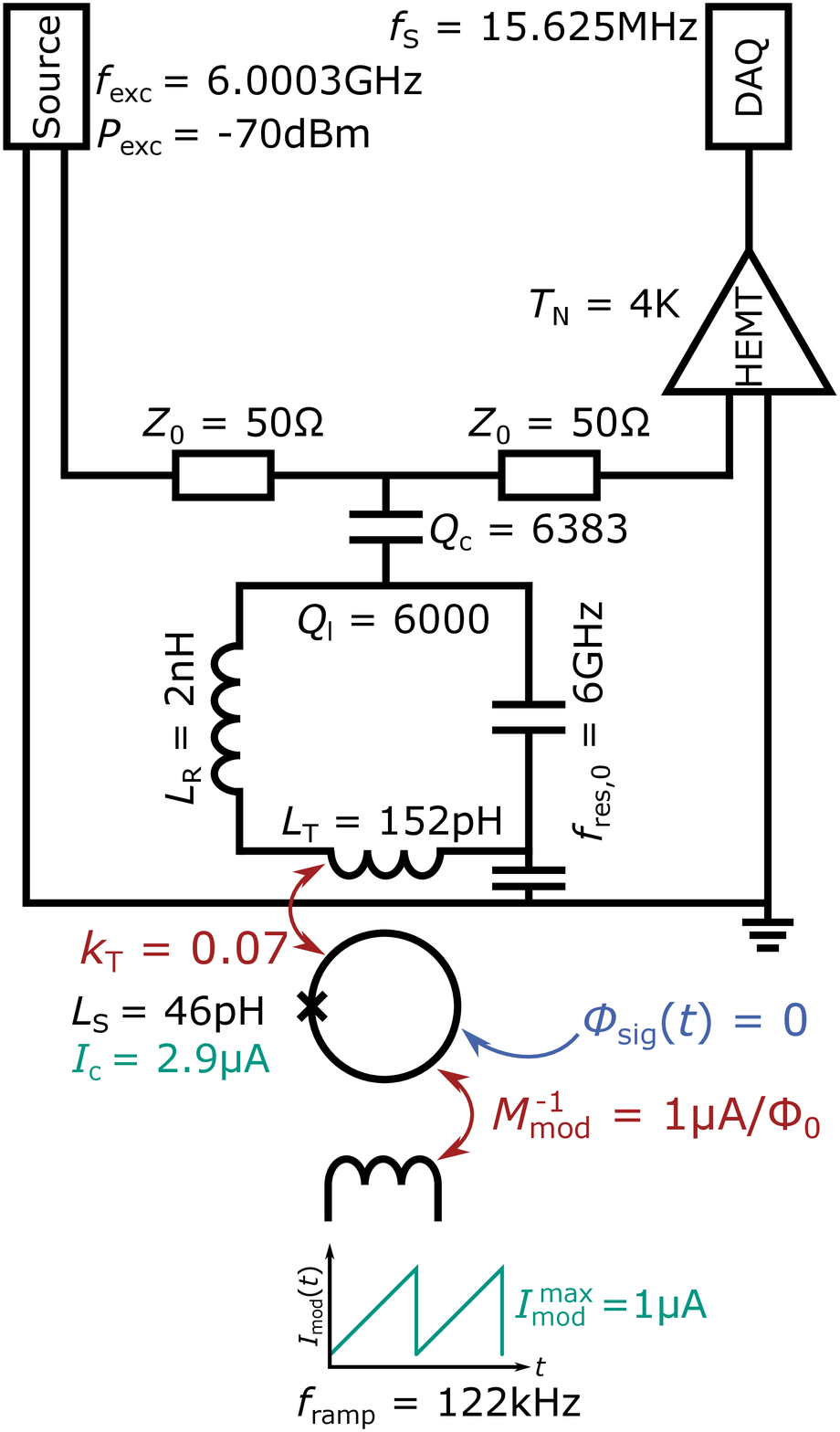}
  \caption{Schematic circuit diagram of a single readout channel as considered during numerical simulation. Non-zero default values of device- and readout parameters are depicted. The different parameters are explained in the main text. The odd value of the coupling quality factor $\Qc$ was chosen to yield a loaded quality factor $\Ql = 6000$ assuming an internal quality factor of $\Qi = \num{1e5}$. Parameters set to zero such as noise sources not used by default are omitted for clarity.}
  \label{fig:default_params}
\end{figure}

In the remaining part of the paper, we present some sanity checks and compare simulation results to experimental data to prove the reliability of our simulation framework. Furthermore, we discuss predictions of our simulation framework aiming towards a full multiplexer optimizations. As the number of input parameters is fairly large (see figure~\ref{fig:numsim_schem}), an enormous number of simulations would be required to perform a full multiplexer optimization. For this reason, we start with varying only a small subset of simulation parameters and postpone the discussion of a full multiplexer characterization to a later publication. The default set of simulation parameters is depicted in figure~\ref{fig:default_params} and is based on our recent activities regarding the development of a microwave SQUID multiplexer for the ECHo experiment \cite{Gastaldo2017} which aims to investigate the electron neutrino mass with sub-eV/$c^2$ sensitivity. 

The multiplexers used for the ECHo experiment employ lumped element microresonators that are formed by a meander-shaped inductor with inductance $\LR = 2\,\mathrm{nH}$, a load inductor with inductance $\LT = \SI{152}{\pH}$ and an interdigital capacitor whose capacitance $C$ is set to yield a unique unloaded resonance frequency $\fresO$ in the frequency band from $\SI{4}{\GHz}$ to $\SI{8}{\GHz}$. For our simulation, we set $\fresO = \SI{6}{\GHz}$ if not otherwise noted. We assume an internal quality factor of $\Qi = \num{1e5}$ as typically measured for our resonators and adjust the effective coupling capacitance $\CCeff$ (including both, the coupling inductance $\CC$ and the parasitic inductance $\Cpar$, see section \ref{sec:intro}) to yield a bandwidth of $\dfBW = \SI{1}{\MHz}$. Moreover, we set the SQUID loop inductance to $\LS \simeq \SI{46}{\pH}$ and adjust the critical current $\Ic$ of the Josephson tunnel junction to yield a screening parameter $\betaL =  0.4$, unless noted otherwise. The mutual inductance $\MT$ is usually chosen to yield $\dfresmax = \dfBW$ and tuned by changing the value of the coupling factor $\kT$. 
It is worth noting that the Bessel function of the first kind, which appears in equation~\ref{eq:dLt_general} can lead to jumps in the SQUID response if the radio frequency flux amplitude becomes too large. To mitigate this effect, the value of the coupling factor $\kT$ has been restricted such that the resonance frequency shift does not exceed 5 times the resonator bandwidth $\dfresmax \leq 5\dfBW = \SI{5}{\MHz}$. For different values of the internal quality factor $\Qi$, this threshold may have to be adjusted.
We assume a sampling rate of $\fs=\SI{15.625}{\MHz}$ that corresponds to the effective sampling rate of the DAQ system presently developed for the ECHo experiment \cite{Wegner2018, Karcher2020}.

For flux ramp modulation, we select a default ramp height of $1\,\PhiO$ at a ramp reset rate of $\framp = \fs/128 \approx \SI{122.1}{\kHz}$ with infinitely fast resets. This rate is low enough to exclude noise degradation due to the finite resonator response time. Additionally, this yields ramp segments with $2^7 = 128$ data points each such that our FFT algorithms works without zero padding. The excitation frequency is assumed to be $\fexc = \fresO + \SI{0.3}{\MHz}$ and is hence slightly above the largest resonance frequency $\fres$ reached during modulation. Finally, we consider, unless noted otherwise, only amplifier white noise with an effective input noise temperature of $T_\mathrm{N} = \SI{4}{\K}$ as resulting from state-of-the-art HEMT amplifiers connected to the multiplexer via superconducting coaxial cables and a cryogenic isolator. Finally, it is worth to mention that we use the most general multiplexer model, i.e. $\betaL > 0, \phirf > 0$, for all simulations discussed in the following.
\section{Validation of the simulation framework}
In order to verify that our simulation framework works as intended, we have performed a number of sanity checks. We verified all functions and simulation steps within a single simulation run including noise generation, flux ramp modulation and demodulation as well as the calculation of the transmission coefficient $\Stok(t)$. Moreover, we comprehensively compared simulation results to experimental data and verified consistency among related data sets. While most of these tests are quite basic and hence not appropriate to be discussed within a paper, we want to discuss two somehow more advanced examples. 
They not only prove the correct functionality of our simulation framework, but also impressively show that our software is able to describe experimental data which is hard to describe otherwise.

\subsection{Dependence of the flux noise on the probe tone power} \label{sec:experimental_comparsion}
\begin{figure}
  \includegraphics[width=0.8\columnwidth]{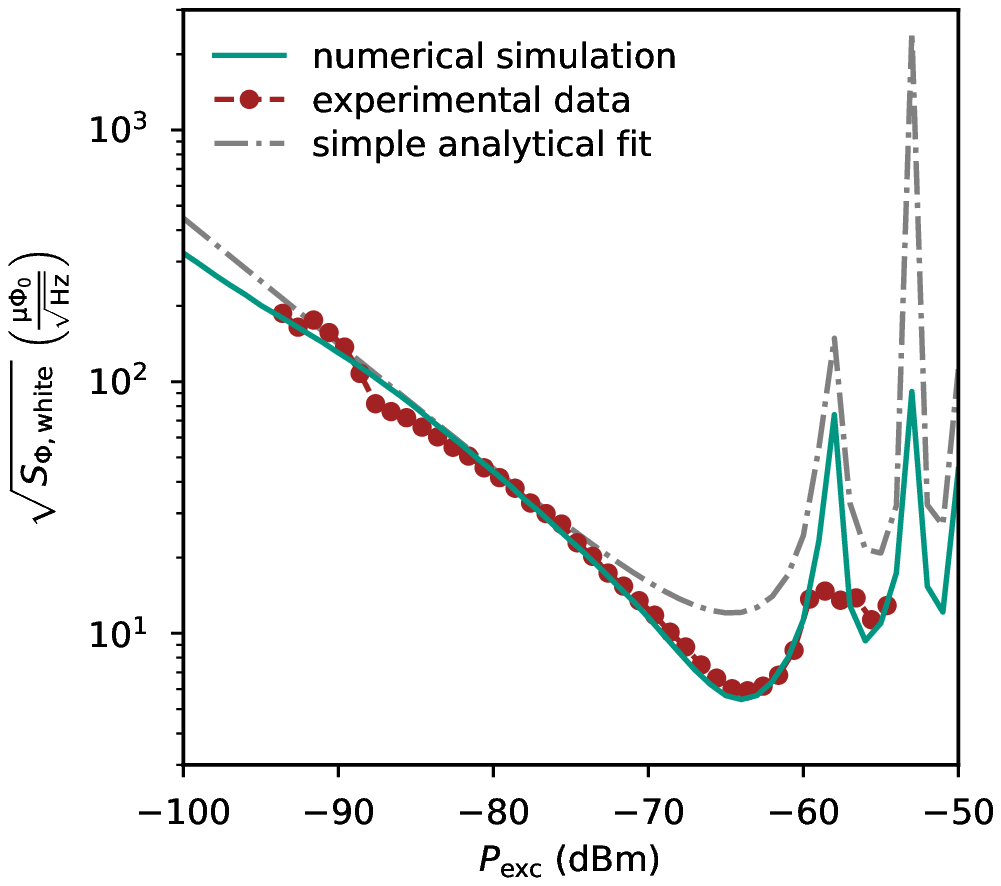}
  \caption{Measured dependence of the white magnetic flux noise level $\sqrt{\Sphiw}$ on the probe tone power $\Pexc$. The data were acquired for an example channel of our most recent microwave SQUID multiplexer with lumped-element microresonators. Details about the device and readout parameters are given in the main text. The measurement was done in open-loop mode, i.e. without flux ramp modulation. In addition to measured data, the expected dependence using a simple analytical approach as well as simulation results as obtained with our simulation framework are shown.}
  \label{fig:SPhiwcomparison}
\end{figure}

The evaluation of the noise performance of a microwave SQUID multiplexer for a given set of device and readout parameters is one of the core applications of our simulation framework. As such, a comparison between experimental data and simulation results of the dependence of the square root of the white magnetic flux noise density $\sqrt{\Sphiw}$ on probe tone power $\Pexc$ provides a reasonable sanity check. To make such a comparison, we comprehensively characterized one of our most recent microwave SQUID multiplexers based on lumped element microresonators and compared the acquired data to simulation results. Figure~\ref{fig:SPhiwcomparison} shows both, measured data and simulation results, of an example multiplexer channel having an unloaded resonance frequency of $\fresO = \SI{4.86}{\GHz}$, while the resonator bandwidth $\dfBW$, maximum resonance frequency shift $\dfresmax$, internal quality factor $\Qi$ and screening parameter $\betaL$ take values of $\dfBW = \SI{3.1}{\MHz}$, $\dfresmax = \SI{0.95}{\MHz}$, $\Qi = 6400$ and $\betaL = 0.4$, respectively. The measurement was performed with open-loop readout at a fixed magnetic bias flux $\Phibias \approx \SI{0.25}{\Phi_0}$. Using the measured or predefined multiplexer and readout parameters, we afterwards simulated the expected dependence $\sqrt{\Sphiw}(\Pexc)$ using our simulation framework. The only free parameter in the simulation was the effective system noise temperature $\TN$ which we haven't determined experimentally. The agreement between experimental data and simulation results is quite impressive in particular close to and below the minimum where the multiplexer would be operated in a real application. The only slight deviation is close to the peak around $\SI{-60}{\dBm}$ and results from the measurement uncertainty of the multiplexer transfer coefficient which gets very small close to the peaks in the flux noise spectral density (we refer the interested reader to \cite{Wegner2022} for a detailed discussion of the reason of the peak occurrence). Overall, this nicely proves that our simulation framework is able to reproduce the characteristics and performance of real multiplexer devices.

It is worth mentioning that one might be inclined to describe the dependence of the measured white noise level on probe tone power directly using the expression as given by our most recent multiplexer model \cite{Wegner2022}. In this case, we expect that the experimental data should follow the dependence $\sqrt{\Sphiw(\phirf)} \propto J_1^{-1}(\phirf)$ ("simple analytical fit" in figure~\ref{fig:SPhiwcomparison}). For low values of the excitation power $\Pexc$, the dependence should follow the intuitive expectation $\sqrt{\Sphiw} \propto 1/ \sqrt{\Pexc}$. However, as the probe tone power $\Pexc$ and thus the rf magnetic flux $\Phirf$ within the SQUID loop increases, the amplitude of the SQUID response is expected to decrease \cite{Wegner2022}. This leads to a degradation of the signal-to-noise ratio in the transmission $\Sto(t)$ and thus to an increase of noise after flux ramp demodulation. Ultimately, this results in a distinct global minimum, the position and depth of which depend on the device parameters, as well as in an oscillatory behavior for large probe tone powers. Here, the dependence of the SQUID response on probe tone power dominates the behavior of the flux noise, and wherever the SQUID response amplitude vanishes, a sharp peak in the readout flux noise occurs. 

While this simple expectation gives a rough estimate for the dependence of the white noise level of the square root of the magnetic flux noise spectral density $\sqrt{\Sphiw}$ on the probe tone power $\Pexc$, it can clearly be seen that the shape of the minimum is not described well. On the other hand, the simulated data models the experimental results much more closely, especially around the minimum. For the experimental data, the region for low readout powers is slightly convex rather than strictly linear as in the analytical description, leading to a lower minimum of a slightly different shape. Since the SQUID response for non-zero screening currents is no longer sinusoidal, the optimal value magnetic bias flux $\Phibiasopt$ depends on the probe tone power $\Pexc$. If a constant value $\Phibias$ of the bias flux is used, as done in the measurement depicted here, this leads to an additional factor influencing the total readout flux noise. While a simple analytical model does not include this contribution, it is described with excellent agreement by the simulations. 

\subsection{Bandwidth- and noise penalty of hybrid microwave SQUID multiplexing}
\begin{figure}
  \includegraphics[width=0.8\columnwidth]{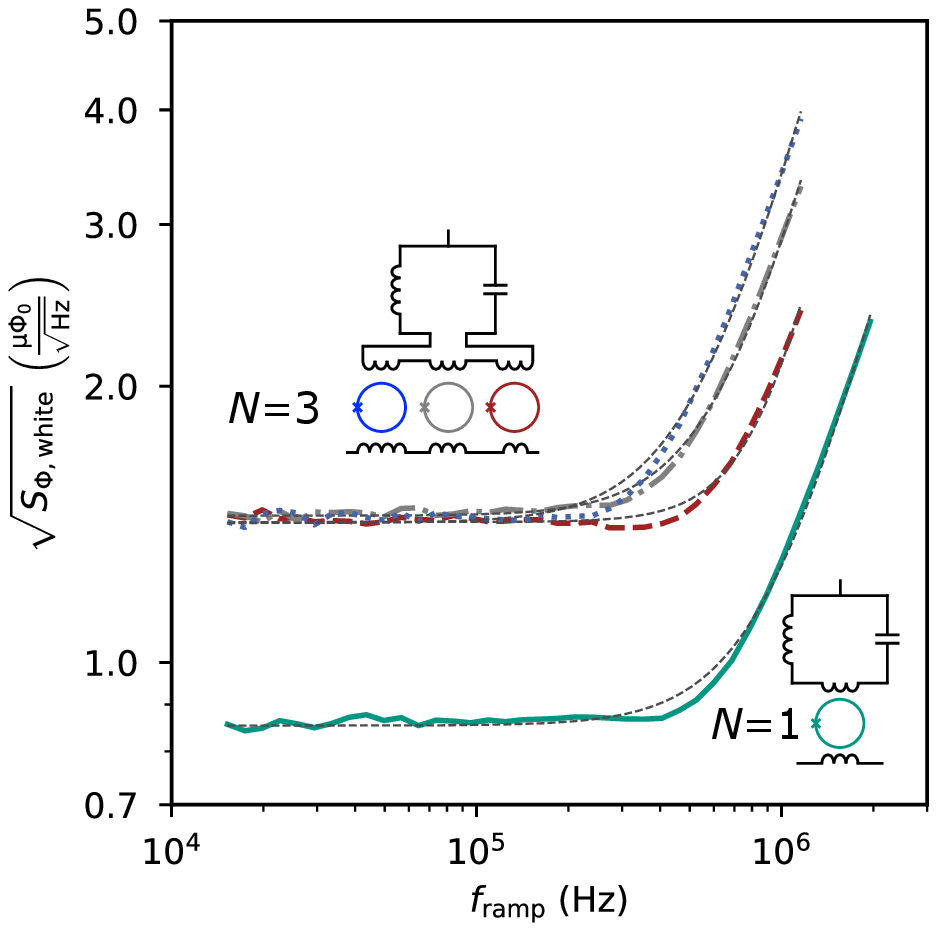}
  \caption{Dependence of the white magnetic flux noise level  $\sqrt{\Sphiw}$ on the flux ramp reset rate for a hybrid microwave SQUID multiplexer with $N=1$ and $N=3$. The inset shows the respective {\HyMUX} configuration where the colors correspond to each other. For further analysis, an empirical function was fitted to each curve (thin grey dashed lines, see main text). 
  }
  \label{fig:HyMUX_framp}
\end{figure}

Flux ramp modulation based hybrid SQUID multiplexing is a very recent multiplexing scheme that allows reducing the number of readout resonators within a microwave SQUID multiplexer while keeping the number of readout channel constant \cite{HyMUX2022}. It might revolutionize SQUID based multiplexing of large-scale bolometric cryogenic detector arrays with a bandwidth in the $\mathrm{kHz}$ range as fabrication accuracy is presently setting strong constraints on the number of readout channels there \cite{HyMUX2022}. A hybrid SQUID multiplexer ({\HyMUX}) closely resembles a regular {\uMUX}, with the major difference that multiple (instead of a single) rf-SQUIDs are coupled to the termination inductance $\LT$ of a readout resonator (see inset of figure~\ref{fig:HyMUX_framp}). Each SQUID is equipped with an individual input coil and coupled to the FRM modulation coil with different strength. During flux ramp modulation, each rf-SQUID experiences a different modulation frequency, transducing the different input signals into unique sidebands of the microwave carrier signal probing the readout resonator. In the subsequent two-step demodulation process, the individual input signals are reconstructed. Because of the strong similarity between both SQUID multiplexer types, our simulation software can also be applied to investigate the properties and characteristics of such an advanced hybrid microwave SQUID multiplexer.

\begin{figure*}
  \includegraphics[width=0.8\textwidth]{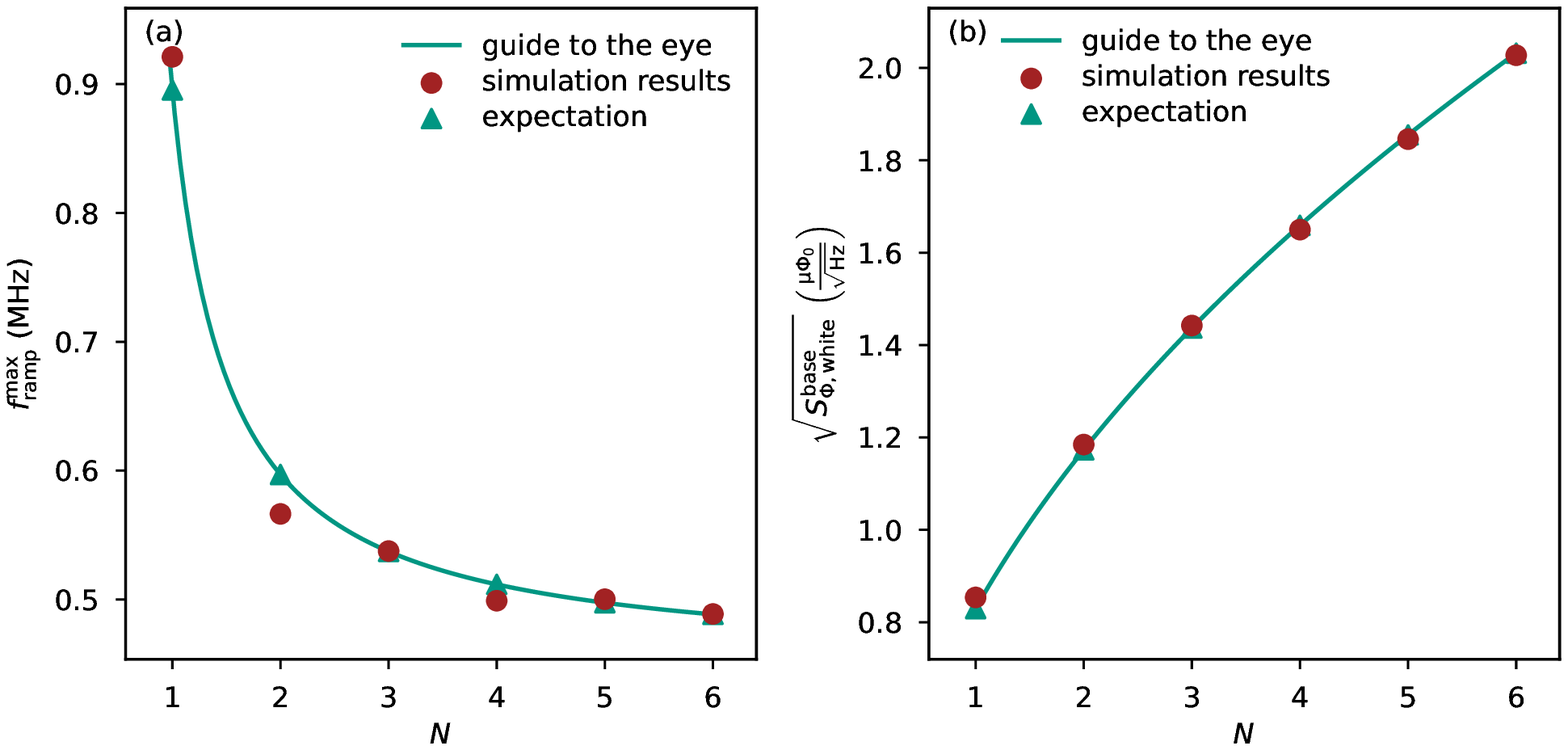}
  \caption{(a) Limit frequency $\frampmax$ of a {\HyMUX} device versus the SQUID number $N$ as extracted from our simulations and as predicted from basic information theory. The SQUID with the lowest limit frequency $\framplim$ limits the channel bandwidth and has been depicted. A guide to the eye for non-integer $N$ is provided as well. (b) Average white noise level $\sqrt{\Sphiwbase}$ in the low frequency limit for each simulated device, as well as predicted from information theory.}
  \label{fig:HyMUX_eval}
\end{figure*}

As another sanity check of our simulation framework, we tried to reproduce the intrinsic bandwidth- and noise penalty of such a hybrid microwave SQUID multiplexer. For this, we performed several simulation runs to determine the dependence of the white noise level of the overall flux noise spectral density $\sqrt{\Sphiw}$ on the flux ramp reset rate $\framp$ for six different {\HyMUX} devices, differing only by the number $N$ of SQUIDs coupled to the resonator as well as their resonator bandwidth $\dfBW$. 
Bandwidth and probe tone power scale linearly with the SQUID number, i.e ${\dfBW}_{,N} \propto N$ and ${\Pexc}_{,N} \propto N$ respectively. For $N=1$, i.e. a conventional microwave SQUID multiplexer, the default values ${\dfBW}_{,1} = \SI{1}{\MHz}$ and ${\Pexc}_{,1} = \SI{-70}{dBm}$ were assumed. Figure~\ref{fig:HyMUX_framp}(a) shows as an example the simulation results for $N=1$ and $N=3$. It clearly shows that at slow flux ramp reset rates $\framp \ll \SI{1}{\MHz}$ the white flux noise level $\sqrt{\Sphiw}$ has a constant base value $\sqrt{\Sphiwbase}$. However, as the ramp reset rate increases, the flux noise level starts increasing above some limit frequency $\framplim$ as the resonator can no longer follow the SQUID modulation due to its finite response time. The limit frequency $\framplim$ takes different values for each SQUID because of the different mutual coupling between SQUID loop and modulation coil, resulting in a variation of modulation frequencies. The maximum flux ramp reset rate suitable for operating the device is ultimately limited by the SQUID with the lowest limit frequency. For determining this frequency, we fitted each curve by the empirical function 
\begin{equation}
    \sqrt{\Sphiw(f)} = \sqrt{\Sphiwbase} \sqrt{ 1 + \left( \frac{f}{\framplim} \right)^b},
\end{equation}
(see figure~\ref{fig:HyMUX_framp}). Considering basic information theory, two relations for {\HyMUX} with a constant number of total readout channels and constant total readout power can be concluded \cite{HyMUX2022}. The maximum usable flux ramp reset rate ${\frampmax} = \mathrm{min}({\framplim} _i)$ (determining the lowest limit frequency of all SQUIDs) as well as the white noise level of each readout channel can be described by
\begin{equation}
    \frampmax \propto \frac{N}{(2N -1)}
\end{equation}
and
\begin{equation}
    \sqrt{\Sphiw} \propto \sqrt{N}.
\end{equation}
Figures~\ref{fig:HyMUX_eval}(a) and (b) show the limit frequency white $\framplim$ as well as white noise level $\sqrt{\Sphiwbase}$ as extracted from our simulations and as predicted by basic information theory. The agreement is excellent and proves that our simulation framework can even describe more complicated multiplexer devices.

\section{First step towards full {\uMUX} optimization}

Our simulation framework allows determining the white noise level $\sqrt{\Sphiw}$ of a single multiplexer readout channel for a predefined set of device and readout parameters within a couple of minutes. This allows finding a parameter configuration which minimizes the overall noise level by systematically varying different parameters. Ideally, the entire configuration space is varied within a set of multiple simulation runs to find a fully optimized device. However, the complexity and hence computational time is exponentially increasing with the number of varied parameters (cf. discussion above). For this reason and as the full optimization of a microwave SQUID multiplexer is not within the scope of this paper, we restricted the parameter space to a small subset and discuss as an example the optimization of readout noise on (i) the screening parameter $\betaL$, (ii) probe tone frequency $\fexc$, (iii) the readout flux $\Phirf$ probing the rf-SQUID, and (iv) the value for the ratio $\eta$ between the maximum frequency shift $\dfresmax$ and the resonator bandwidth $\dfBW$.

\subsection{Dependence of readout noise on screening parameter $\betaL$ and probe tone frequency $\fexc$}

\begin{figure*}
\centering
  \includegraphics[width=0.8\textwidth]{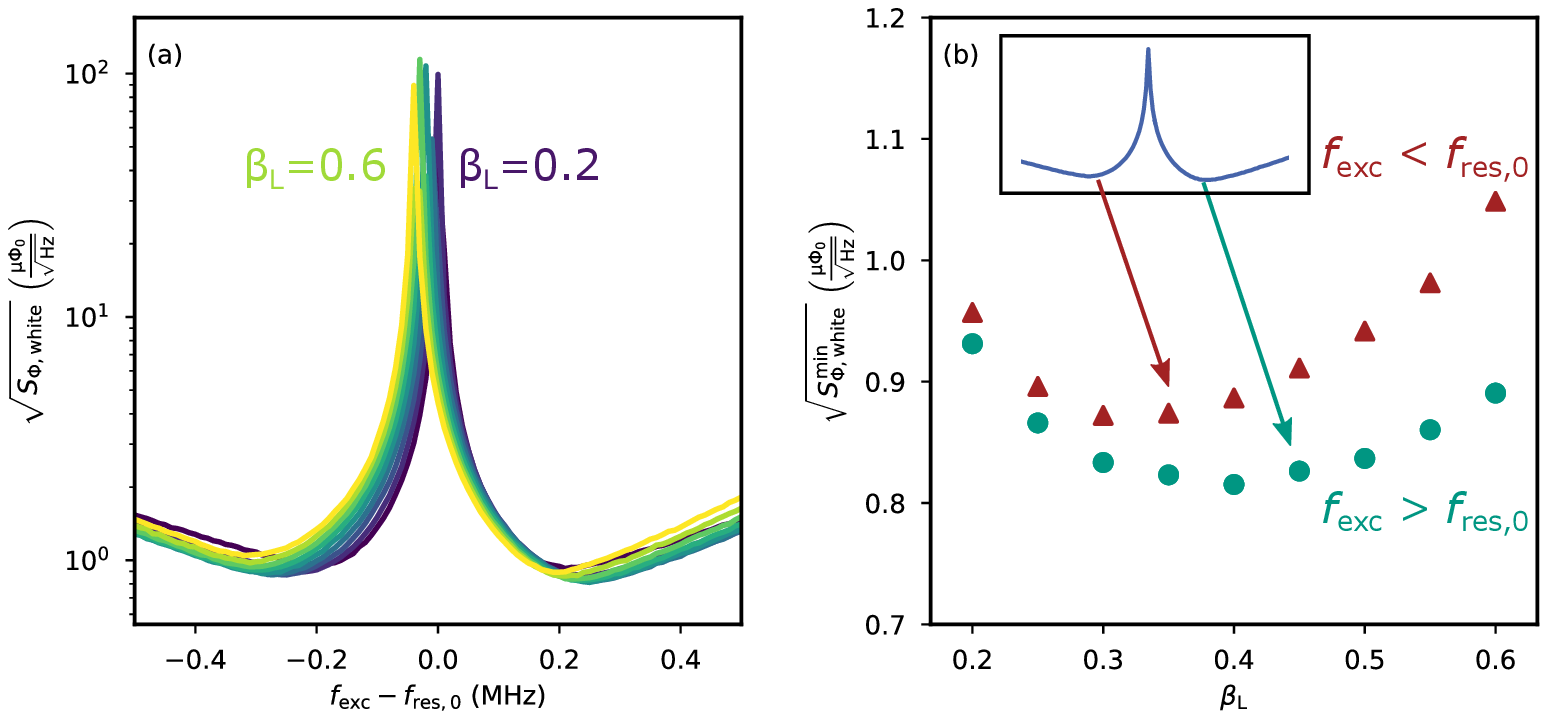}
  \caption{(a) White readout flux noise $\sqrt{\Sphiw}$ as a function of the probe tone frequency $\fexc$ for different values of the SQUID screening parameter $\betaL$. The latter was altered by varying the critical current $\Ic$. (b) Minimum white noise level $\sqrt{\Sphiw^\mathrm{min}}$ in both local minima of figure (a) for each value of $\betaL$.}
  \label{fig:sim_fro}
\end{figure*}

We run a dedicated set of simulations to determine the dependence of the white noise level $\sqrt{\Sphiw}$ on the probe tone frequency $\fexc$ for several values of the screening parameter $\betaL$. For each simulation, the mutual inductance $\MT$ between SQUID and resonator was tuned to guarantee $\dfresmax = \dfBW \SI{1}{\MHz}$. The corresponding results are depicted in figure~\ref{fig:sim_fro}(a). If the probe tone frequency $\fexc$ is very close to the unaltered resonance frequency $\fresO$, i.e. $\fexc - \fresO \approx 0$, the actual resonance frequency switches from being below to being above $\fexc$ during flux ramp modulation. In this scenario, the shape of the resulting transmission response is non-sinusoidal, compromising FRM demodulation and resulting in enhanced readout noise (see appendix~\ref{sec:noise_divergence} for more details). This manifests as the central peak in the figure. To either side of the central peak, a local minimum is found. The asymmetry of the curves is related to the asymmetry of the SQUID response due to the non-linear junction equations \cite{Wegner2022}.

In figure~\ref{fig:sim_fro}(b), we show the white noise values of both minima for $\fexc > \fresO$ and $\fexc < \fresO$, respectively, for several values of the SQUID screening parameter $\betaL$. It is obvious that for any choice of the screening parameter $\betaL$, choosing $\fexc > \fresO$ yields a lower overall white noise level. This agrees well with our expectation regarding the dependence of both, the resonator transmission spectrum and the SQUID response, on the external magnetic flux. For non-zero values of the screening parameter $\betaL$, the latter is non-sinusoidal and is further distorted when transduced to a transmission response by the resonator. With $\fexc > \fresO$, the resulting transmission response is closer to a sinusoidal shape, thus leading to a more efficient demodulation and lower readout noise. The broad minimum for $0.3 < \betaL < 0.5$ shows that for a given parameter set the noise level does not strongly depend on $\betaL$. Assuming the mutual inductance $\MT$ can be tuned in a post fabrication process to yield $\dfresmax = \dfBW$, this significantly relaxes junction fabrication as $\Ic$ can easily vary due to fabrication inaccuracies. In figure~\ref{fig:sim_phirf}(b) (red), the probe tone frequency yielding the lowest overall noise performance is shown as a function of the screening parameter $\betaL$. It nicely shows that the optimal excitation frequency $\fexc^\mathrm{opt}$ can be determined for any set of device parameters.

\subsection{Optimal value of rf flux amplitude $\Phirf$ within the SQUID loop}

A critical parameter for {\uMUX} operation is the rf magnetic flux amplitude $\Phirf$ used for probing/exciting the SQUID. It is set by the probe tone power as well as several design parameters such as the SQUID screening parameter $\betaL$ and strongly affects the system white noise level for amplifier limited setups. To investigate the complex interplay between the associated device and readout parameters, the dependence of the white readout flux noise $\sqrt{\Sphiw}$ on the probe tone power $\Pexc$ was simulated for different values of the screening parameter $\betaL$. The mutual inductance $\MT$ was adjusted such that $\dfresmax = \dfBW = 1\,\mathrm{MHz}$.

\begin{figure*}
\centering
  \includegraphics[width=0.8\textwidth]{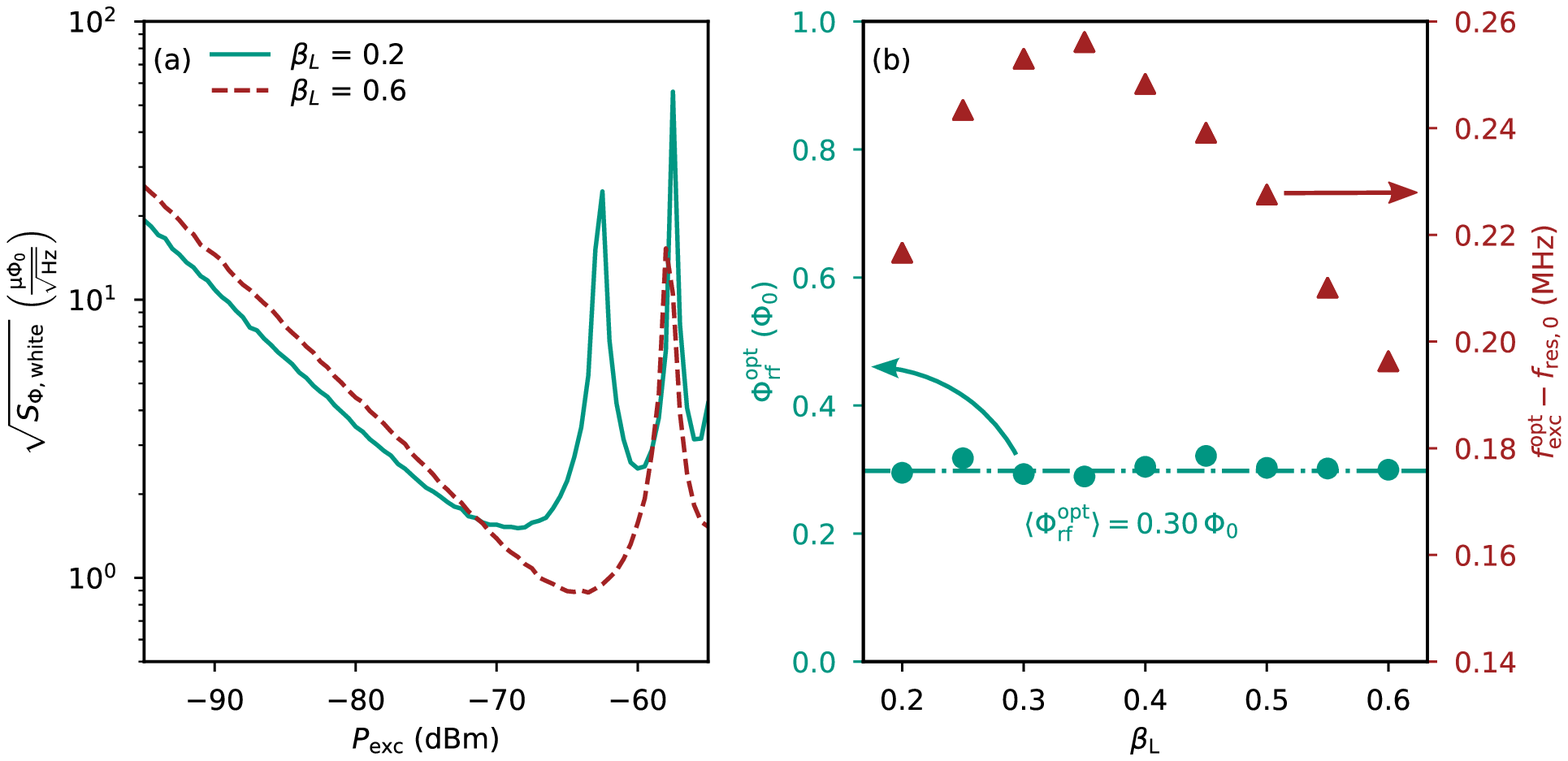}
  \caption{(a) White readout flux noise $\sqrt{\Sphiw}$ as a function of the probe tone power $\Pexc$ for two different values of $\betaL$. (b) Optimum rf magnetic flux amplitude $\Phirf^\mathrm{opt}$ yielding the lowest noise floor ($\bullet$, left y-axis) and optimum probe tone frequency $\fexc^\mathrm{opt}$ ($\blacktriangle$, right y-axis) versus the SQUID screening parameter $\betaL$. }
  \label{fig:sim_phirf}
\end{figure*}

Figure~\ref{fig:sim_phirf}(a) shows as an example two acquired simulation curves.
The overall shape of the resulting curves was already discussed in section~\ref{sec:experimental_comparsion}. At low excitation powers $\Pexc$, the dependence of the white readout flux noise $\sqrt{\Sphiw}$ on the probe tone power results from the increasing signal-to-noise ratio, yielding a linear decrease of noise level. At high excitation powers $\Pexc$, the power dependence of the SQUID response dominates, resulting in an oscillatory behavior. In between, a distinct global minimum forms, the position and depth of which depend on the value of the screening parameter $\betaL$.

In figure~\ref{fig:sim_phirf}(b), we show the dependence of the radio frequency magnetic flux amplitude $\Phirf^\mathrm{opt}$ at the probe tone power which minimizes readout flux noise on the screening parameter $\betaL$.
It is apparent that $\Phirf^\mathrm{opt}$ is independent of $\betaL$ and that the ideal value is $\Phirf^\mathrm{opt} \approx 0.30\,\PhiO$. Similar results are obtained for other resonance frequencies $\fresO$ indicating a universal behavior. This observation is in good agreement with results reported in \cite{Mates2017}.

\subsection{Optimal ratio between maximum frequency shift and the resonator bandwidth}

\begin{figure*}
\centering
  \includegraphics[width=0.8\textwidth]{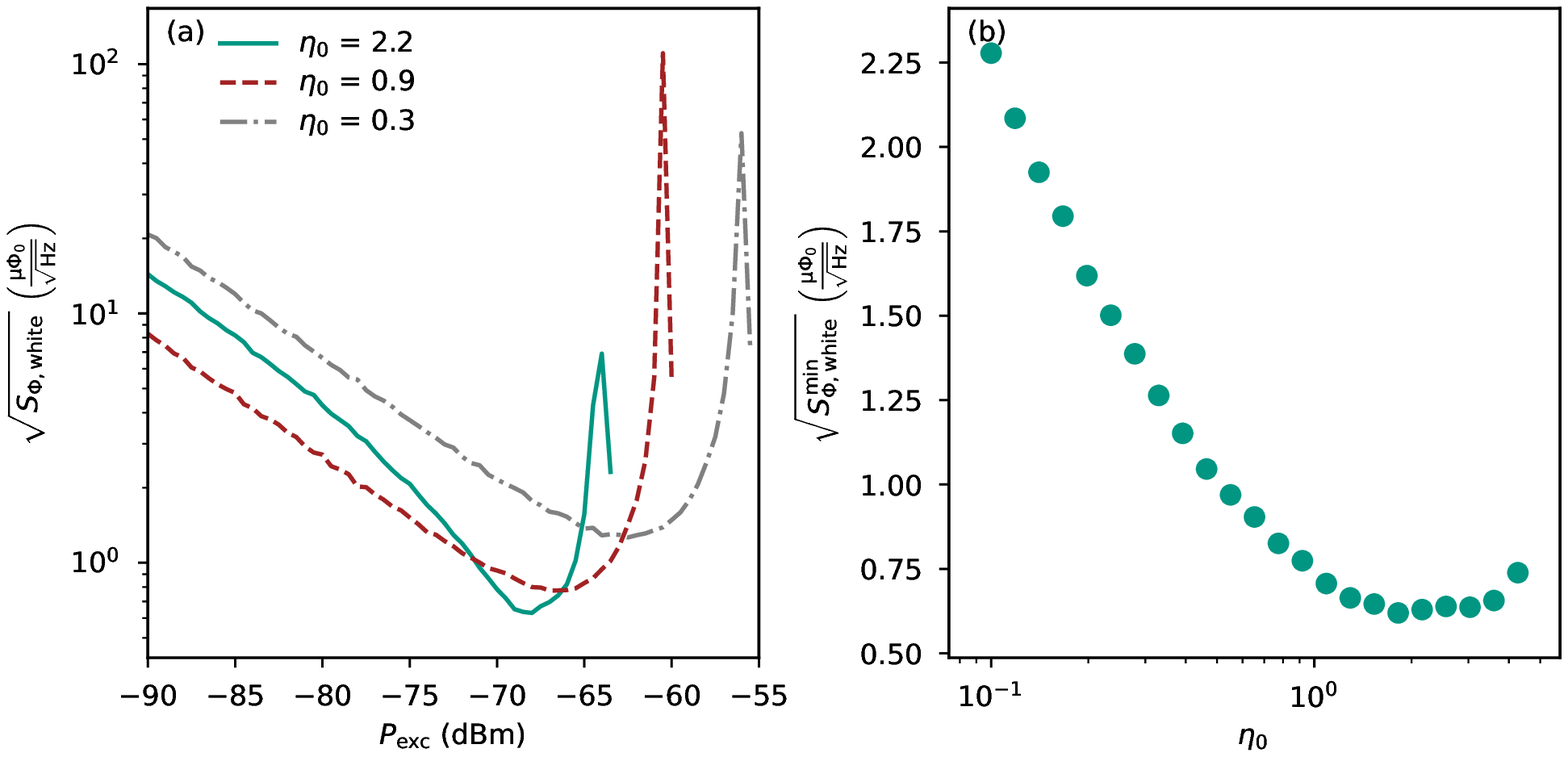}
  \caption{(a) White noise level $\sqrt{\Sphiw}$ of the magnetic flux as a function of the microwave probe tone power $\Pexc$ for three values of $\eta_0$, set by changing the coupling mutual inductance $\MT$. (b) Minimum noise value $\sqrt{\Sphiw^\mathrm{min}}$ versus the ratio parameter $\eta_0 = \lim_{\Pexc \to 0} \etaeff(\Pexc)$.}
  \label{fig:eta_result}
\end{figure*}

The ratio $\eta = \dfresmax/\dfBW$ between the maximum frequency shift $\dfresmax$ and the resonator bandwidth $\dfBW$ is usually chosen close to unity, i.e. $\eta \approx 1$, to guarantee optimal readout conditions \cite{Kempf2017,Mates2017}. 
However, this rule of thumb does not take into account that $\dfresmax$, and hence the ratio $\eta$, both depend on the probe tone power used for resonator readout, i.e. $\dfresmax = \dfresmax(\Pexc)$ and $\eta = \eta(\Pexc) \dfresmax(\Pexc)/\dfBW$. Since the maximum frequency shift and the overall white noise floor both depend on the probe tone power, we have to expect a severe deviation from the empirical value $\eta \approx 1$. For this reason, we investigated the effect of $\eta$ on the readout noise. We performed a set of simulations for which we systematically varied the probe tone power $\Pexc$ for various values of the low-power value $\eta_0 = \lim_{\Pexc \to 0} \eta(\Pexc)$.
Figure~\ref{fig:eta_result} summarizes the results of these simulations.
Figure~\ref{fig:eta_result}(a) nicely shows that the position of the noise minimum shifts towards lower readout power as $\eta_0$ increases due to the related increase in coupling mutual inductance $\MT$. In the limit of low probe tone powers $\Pexc \to 0$, the typical choice of $\eta_0 \approx 1$ indeed leads to the best noise level. However, as $\Pexc$ approaches its ideal value, the overall minimal readout noise is achieved for $\eta_0 > 1$. Moreover, figure~\ref{fig:eta_result}(b) shows the dependence of noise in the minimum $\Sphiw^\mathrm{min}$ on the ratio $\eta_0$. It is obvious that with increasing power, $\eta_0 > 1$ turns out to ultimately yield lower noise as compared to the empirical value $\eta_0 \approx 1$. As the resonance frequency modulation amplitude $\dfresmax$ decreases with increasing readout power, so does $\eta(\Pexc)$. A proper choice of $\eta_0$ leads to the ideal value of $\eta(\Pexc^\mathrm{opt}) \approx 1$ at the ideal probe tone power rather than in the low power limit, ultimately decreasing readout noise and reducing $\Pexc^\mathrm{opt}$. The latter is an important result taking into account that present multiplexers somehow suffer from intermodulation products related to the IIP3 points of the subsequent amplifier chain \cite{Henderson2018}. 
Hence, reaching the optimum noise level at lower readout power allows to increase the multiplexing factor for a given amplifier chain.
\section{Potential other applications of the simulation framework}
The main application of our simulation framework is the analysis and optimization of the characteristics and performance of a microwave SQUID multiplexer and associated readout devices. However, it can be additionally used to study a variety of effects related to {\uMUX} operation, two of which we will showcase in this section.

\subsection{Linearity of microwave SQUID multiplexers}
Because of the periodicity of the SQUID response, the output signal of a {\uMUX} has to be linearized.
This is typically achieved using flux ramp modulation \cite{Mates2012}.
However, the complex interplay between the nonlinear characteristics of a microwave SQUID multiplexer and flux ramp modulation make the analysis of output signal linearity in {\uMUX} based readout systems by analytical means unfeasible. Second-order effects such as the finite reset time of the flux ramp and the finite resonator response time add even more complexity. Our simulation framework allows predicting / investigating device linearity in just a few minutes.

\begin{figure*}
  \includegraphics[width=0.8\textwidth]{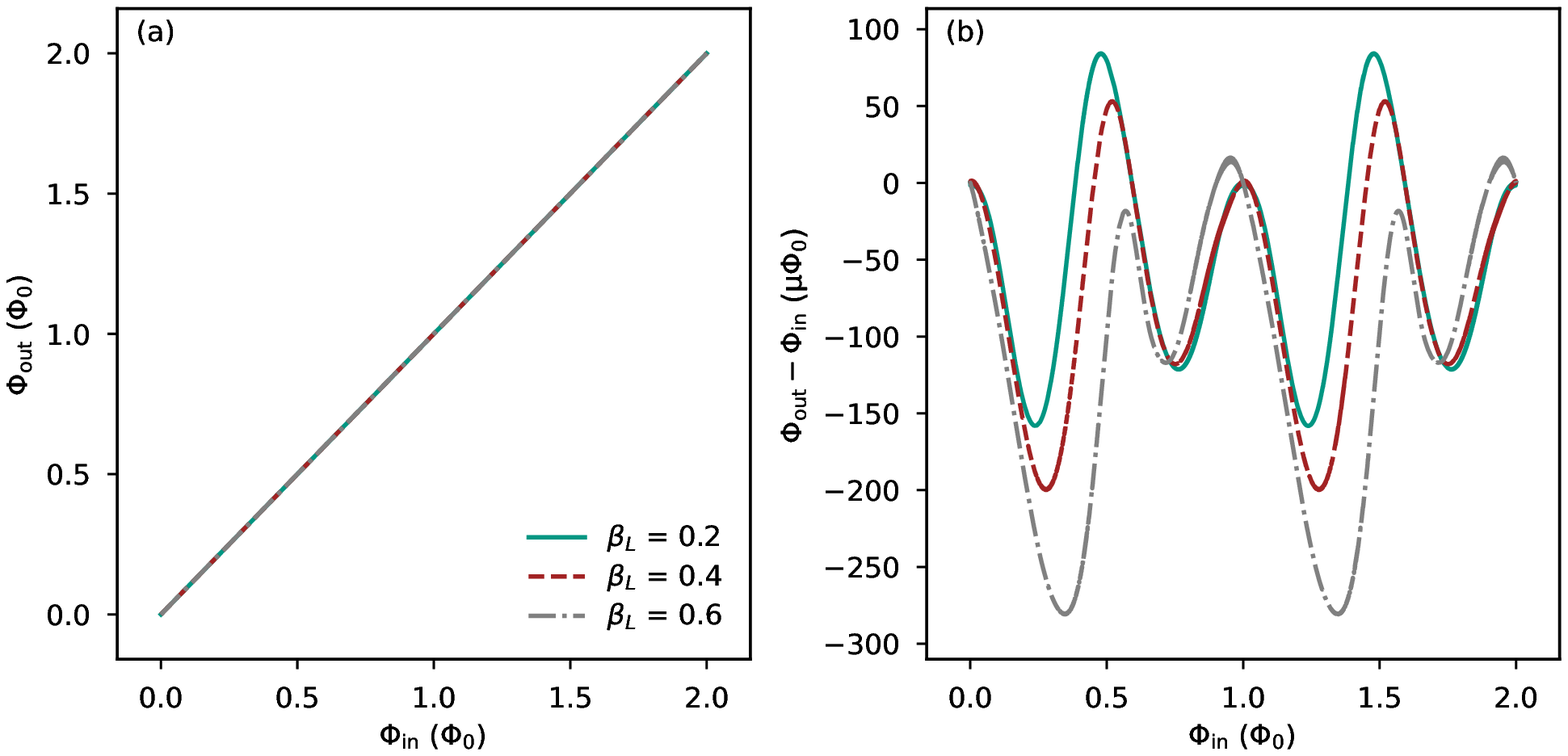}
  \caption{(a) Output signal flux $\Phiout$ as a function of the input signal flux $\Phiin$ for different values of the screening parameter $\betaL$. (b) Deviation from perfect linearity, i.e. the difference $\Phiout - \Phiin$ as a function of the input signal flux $\Phiin$.}
  \label{fig:results_linearity}
\end{figure*}

Figure~\ref{fig:results_linearity}(a) shows as an example the dependence of the output signal flux $\Phiout$ on the input signal flux $\Phiin$ for three microwave SQUID multiplexers with different values of the screening parameter $\betaL$. From a bird's eye view, the relation between input and output signal looks almost ideally linear. However, subtracting a linear fit from the simulated input-output relation reveals a remaining non-linearity which is shown in figure~\ref{fig:results_linearity}(b). For these simulations, a flux ramp with an amplitude of at most $2.5\,\Phi_0$ in the SQUID loop was assumed. A second-order Butterworth low-pass filter with a cutoff frequency of $f_\mathrm{cutoff} = 10\,\mathrm{MHz}$ was applied to emulate a finite flux ramp reset time. Datapoints amounting to $20\%$ of each ramp segment were neglected to avoid transients of the ramp resets to affect the demodulation procedure. We didn't add noise traces to focus on systematic nonlinearity only.

Since the FRM method is based on the phase determination of a periodic signal, its non-linearity also has to be periodic with the same period length of one flux quantum $\PhiO$. This behavior is clearly apparent from figure~\ref{fig:results_linearity}(b). The peak to peak range of the non-linearity is roughly $250\,\mathrm{\mu \PhiO}$ for all three simulated devices. While the value of the screening parameter changes the shape of both the SQUID response and the non-linearity curve, it has very little effect on the magnitude of the deviation from the desired linear behavior for the parameter set assumed here.

\subsection{Noise shaping}
Compared to open-loop readout, flux ramp modulation has a significant effect on the shape of the magnetic flux noise spectral density. At high frequencies, the latter is typically dominated by white noise caused by cryogenic amplifiers. As flux ramp modulation leads to a reduced effective flux-to-transmission transfer coefficient $\Kphi(\Phi)$ as compared to open-loop readout, the white noise level is in general increased by a factor $\cdeg > \sqrt{2}$ when using flux ramp modulation \cite{Mates2012}. At low frequencies a $1/f^\alpha-$like noise contribution due to two-level systems in the vicinity of the microwave resonator has been observed \cite{Gao2007} dominating the overall noise spectral density at low frequency. Due to the nature of the flux ramp modulation, an offset of the transmission signal constant on timescales of the ramp reset rate $\framp$ has no effect on the extracted phase. As a result, TLS noise contributions at frequencies of the order of the ramp reset rate $\framp$ and below do (to first order) not contribute to the overall readout flux noise. Thus, flux ramp modulation can, at least partially, suppress the $1/f^\alpha-$like noise contribution due to TLS. Despite the mathematical complexity of {\uMUX} and the FRM method, the simulation framework allows for a detailed prediction of the resulting readout flux noise spectral density. 

\begin{figure}
  \includegraphics[width=0.8\columnwidth]{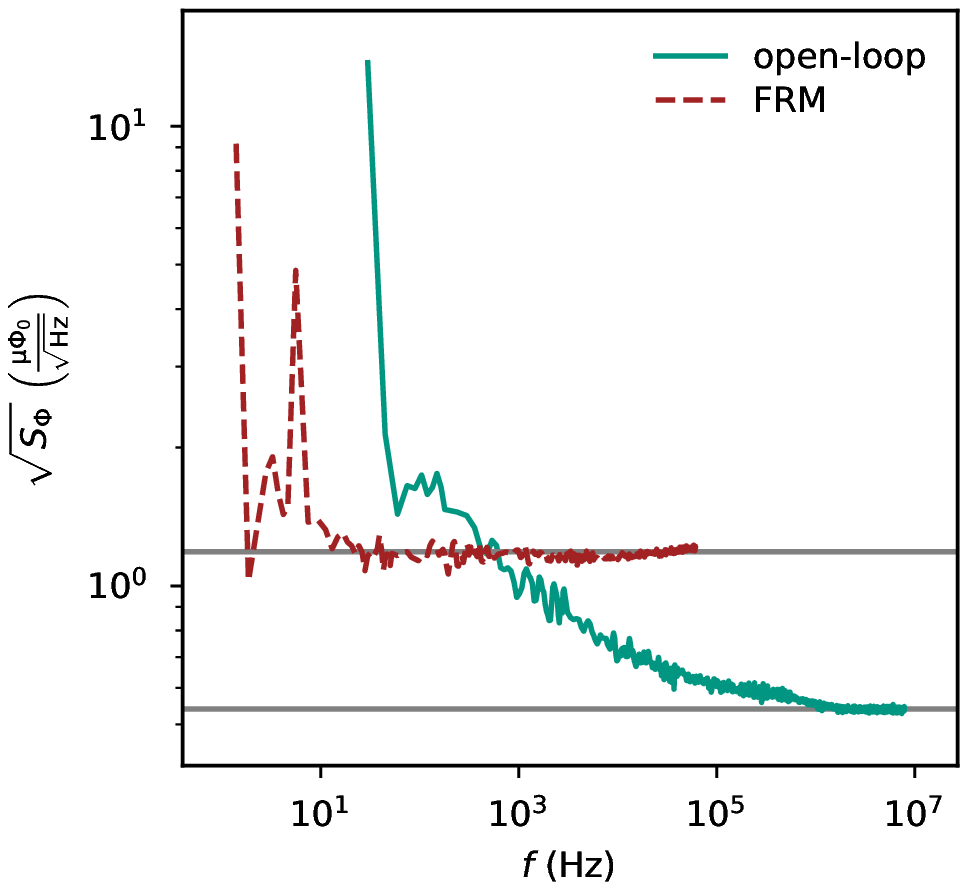}
  \caption{Square root of the flux noise spectral density $\sqrt{\Sphi(f)}$ for both open-loop and FRM readout of the same {\uMUX} device. Grey lines mark the white noise level at high frequencies.}
  \label{fig:results_noiseshaping}
\end{figure}

Figure~\ref{fig:results_noiseshaping} shows the square root of the noise spectral density $\sqrt{\Sphi(f)}$ of an example microwave SQUID multiplexer assuming open-loop and FRM readout. Both, a white amplifier noise with a noise temperature of $\TN = \SI{4}{\kelvin}$ and a $1/\sqrt{f}-$like TLS noise with a noise level of $\sqrt{S_\mathrm{TLS}}/\fresO = \SI{2.5e-9}{1\per\sqrt{\hertz}}$ at a frequency of $\SI{1}{\hertz}$ were assumed for the simulations. These values serve as rough exemplary values for microresonators \cite{Gao2007} \footnote{We are aware that the TLS noise power depends on both probe tone power and resonator geometry, and chose these values purely to showcase the functionality of the simulation software. In a future iteration of the simulation framework, a full model for TLS noise may be included.}. 
For open-loop readout, the bias flux $\Phibias$ was chosen such that the flux-to-transmission transfer coefficient $\Kphi(\Phibias)$ is maximized. For FRM readout, a modulation ramp inducing at most one flux quantum into the SQUID loop with a ramp repetition rate of $\framp = \fs / 128 \approx \SI{122}{\kHz}$ was used. 
Due to the reduction in output sampling rate caused by the demodulation during FRM readout, the square root of the noise spectral density of open-loop readout extends to higher frequencies. At high frequencies where the curves are flat, FRM leads to a higher level of noise compared to open-loop readout by a factor of $\cdeg = 2.18$. This is caused by the reduced effective gain as discussed before. In the case of open-loop readout, the noise contribution due to two-level systems results in a characteristic increase of the square root of the noise spectral density towards low frequencies. As expected, this increase is significantly less prominent for FRM readout, where the noise spectral density is mostly white. Low-frequency noise added in the signal chain after the SQUID is hence significantly reduced, however due to nonlinearities not fully removed. As a result at frequencies around $\SI{10}{\Hz}$ and below, a slight increase of the noise level towards low frequencies is visible even for FRM readout. Similar analysis can help to further investigate the effects that readout schemes like FRM have on the noise of {\uMUX} and may ultimately lead to an improved understanding of the intricate behavior of such devices.
\section{Conclusion}
We presented a software framework to simulate the characteristics and performance of a single channel of a microwave SQUID multiplexer. Our simulation framework is based on the state-of-the-art multiplexer model including a full description of the dependence of device performance on the screening parameter $\betaL$ and the rf flux amplitude $\Phirf$ as well as dynamical effects due to the finite bandwidth of the microwave resonator. Either open-loop or FRM readout can be used for the simulation. To verify that the software works as intended, we performed several tests and showed that our simulation results are in excellent agreement with experimental results. Moreover, we showed that it can describe the expected behavior of more sophisticated devices such as hybrid microwave SQUID multiplexers as long as they are direct derivatives of a conventional {\uMUX}. We presented first steps towards a full optimization of microwave SQUID multiplexers by exploring the dependence of {\uMUX} performance on a small subset of all possible device and readout parameters. We showed, for example, that device performance is better in case that the probe tone frequency is larger than the unloaded resonance frequency $\fexc > \fresO$ and that a value of the SQUID screening parameter in the range $0.3 \leq \betaL \leq 0.5$ yields the minimum magnetic flux noise. Moreover, we showed that the probe tone power $\Pexc$ should be chosen such that a rf flux amplitude $\Phirf = \SI{0.3}{\Phi_0}$ is threading the SQUID loop and that the typical choice of $\eta_0 = 1$ is not universal, and that $\eta_0 > 1$ can result in a lower minimum readout noise at a lower probe tone power. The latter can significantly improve the multiplexing factor in systems limited by the cryogenic amplifier IIP3 point. Finally, we highlighted other applications of our simulation framework such as a discussion of linearity or noise shaping.

\section*{Acknowledgments}
This work was performed within the framework of the DFG research unit FOR 2202 (funding under grant no. En299/7-1 and En299/7-2). C. Schuster further gratefully acknowledges support by the Karlsruhe School of Elementary Particle and Astroparticle Physics: Science and Technology (KSETA).

\section*{Data Availability Statement}
The data that support the findings of this study are available from the corresponding author upon reasonable request.

\appendix
\section{Dynamical lumped element resonator model}
\label{sec:dynresmodel}

\begin{figure}
  \includegraphics[width=0.8\columnwidth]{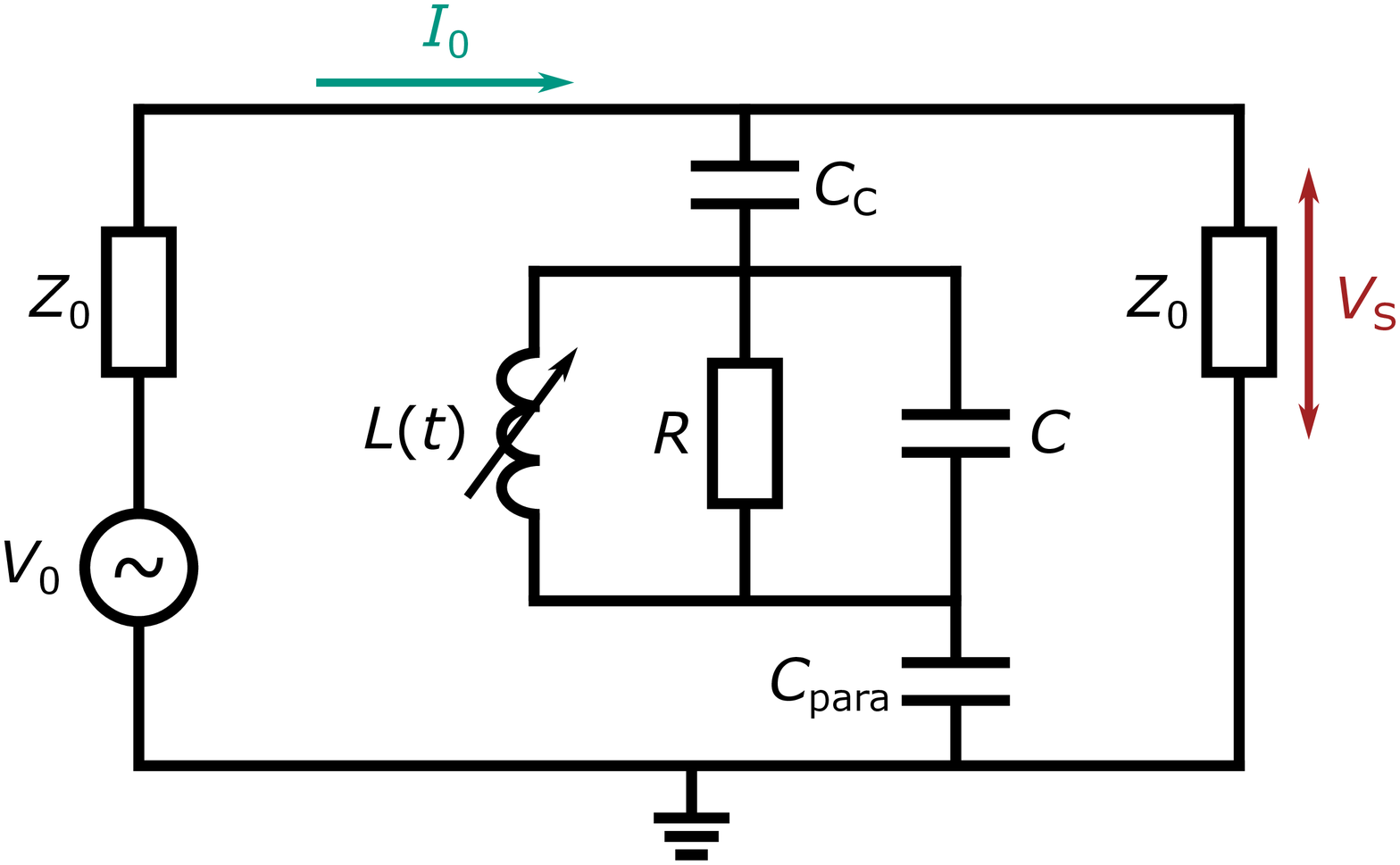}
  \caption{Simplified schematic circuit diagram of a lumped element based microwave resonator with a time-dependent inductance $L$. Parameters are explained in the main text.}
  \label{fig:lemwr_schem_App}
\end{figure}

Figure~\ref{fig:lemwr_schem_App} depicts the schematic circuit diagram of a lumped element microwave resonator consisting of a capacitance $C$, resistance $R$ and time-dependent inductance $L(t)$ connected in parallel. The resistance $R$ represents losses within the resonator. In a {\uMUX} channel, the resonator inductance is modulated by the rf-SQUID, represented by with the time-dependence of the inductance $L$. The resonator is coupled to a transmission line with capacitance $\CC$ and to ground with capacitance $\Cpar$. The transmission line has the impedance $Z_0$. A microwave ac voltage $V_0$ with amplitude $|V_0|$ and angular frequency $\omega$ is applied to its input, causing an ac current $I_0$. The output voltage $\Vs$ is measured to record the transmission parameter $\Sto = 2 \Vs / V_0$. The resonance angular frequency $\omegares$, internal quality factor $\Qi$, coupling quality factor $\Qc$ and resonator bandwidth $\dfBW$ of this configuration are then given by:
\begin{eqnarray} \label{eq:qualitydefs}
\omegares &=& \frac{1}{\sqrt{L \left( \CCeff + C \right)}} = 2 \pi \fres\\
\Qi &=& \frac{R}{\omegares L} \\
\Qc &=& \frac{2}{Z_0 \omegares^3 L \CCeff^2} \\
\dfBW &=& \frac{\fres}{\Ql} \approx \frac{\fres}{\Qc}.
\end{eqnarray}

For the derivation of the transmission coefficient $\Sto$ of a lumped element resonator as shown in figure \ref{fig:lemwr_schem_App}, it is convenient to use the impedances in their operator forms, where we will use the convention $D \equiv \frac{d}{dt}$ for the operator of the time derivative. For inductances and capacitances respectively this yields:
\begin{eqnarray}
Z_L &= LD + \dot{L} \\
Z_C &= \frac{1}{CD}
\end{eqnarray}
The current $I_{0}$ can be derived using Ohm's law and the Kirchhoff rules as follows:
\begin{equation} \label{eq:app_I0}
I_0 = V_0 \left( \frac{P + Z_0 Q}{Z_0 \left( 2 P + Z_0 Q \right) }\right).
\end{equation}
With the substitutions:
\begin{eqnarray}
P &=& D^2 + \frac{1}{R \left( \CCeff + C \right)} \left[ 1 + R \left( \CCeff + C \right) \frac{\dot{L}}{L} \right] D\\
&& + \frac{1}{L \left( \CCeff + C \right)} \left[ 1 + \frac{\dot{L}}{R} \right] \nonumber \\
Q &=& \frac{C_c C }{\CCeff + C} D^3 + \frac{C_c}{R \left( \CCeff + C \right)} \left[ 1 + R C \frac{\dot{L}}{L}\right] D^2\\
&& + \frac{C_c}{L \left( \CCeff + C \right)} \left[ 1 +  \frac{\dot{L}}{R} \right] D \nonumber
\end{eqnarray}

Using equation~\ref{eq:app_I0} and Kirchhoff’s laws we can find:
\begin{equation}
P V_0 = \left(2 P + Z_0 Q \right) V_S
\end{equation}

Using the substitutions

\begin{eqnarray}\label{eq:XYZAB}
X &\equiv \frac{\CCeff C}{\left( \CCeff + C \right)} \quad \quad&A \equiv \frac{1}{R \left( \CCeff + C \right)} \nonumber \\
Y &\equiv \frac{\CCeff}{R \left( \CCeff + C \right)} \quad \quad &B \equiv \frac{1}{L \left( \CCeff + C \right)} \nonumber \\
Z &\equiv \frac{\CCeff}{L \left( \CCeff + C \right)}. \nonumber 
\end{eqnarray}

Using $\Vs = \Sto V_0 / 2$ we can expand this into

\begin{eqnarray}\label{eq:S21hell01}
\left[ Z_0 X D^3 \right. && \\ 
\left. + \left( 2 + Z_0 \left( Y + X \frac{\dot{L}}{L} \right) \right) D^2  \right. && \nonumber \\ 
\left.  + \left( 2  \left( A + \frac{\dot{L}}{L} \right) + Z_0  \left( Z + Y \frac{\dot{L}}{L} \right) \right) D  \right. && \nonumber \\ 
\left. + 2 \left( B + A \frac{\dot{L}}{L} \right) \right] & S_{21} V_0& \nonumber \\ 
= 2 \left[ D^2 + \left( A + \frac{\dot{L}}{L} \right) D + \left( B + A \frac{\dot{L}}{L} \right) \right] V_0. \nonumber
\end{eqnarray}

Applying the differentiation operator and rearranging terms then finally yields the third-order differential equation:

\begin{equation}\label{eq:S21hell02a}
N_4 D^3 \Sto + N_3 D^2 \Sto + N_2 D \Sto + N_1  \Sto = N_0
\end{equation}

with the coefficients

\begin{eqnarray}\label{eq:S21hell02}
N_4 &=& \left[ Z_0 X  \right] \\ 
N_3 &=& \left[ 3 \rmi \omega Z_0 X + \left( 2 + Z_0 \left( Y + X \frac{\dot{L}}{L} \right) \right) \right] \\
N_2 &=& \left[ -3 \omega^2 Z_0 X + 2 \rmi \omega \left( 2 + Z_0 \left( Y + X \frac{\dot{L}}{L} \right) \right) \right. \\
&&+ \left. \left( 2 \left( A + \frac{\dot{L}}{L} \right) + Z_0 \left( Z + Y \frac{\dot{L}}{L} \right) \right) \right] \nonumber \\
N_1 &=& \left[ - \rmi \omega^3 Z_0 X - \omega^2 \left( 2 + Z_0 \left( Y + X \frac{\dot{L}}{L} \right) \right) \right. \\
&&+ \left. \rmi \omega \left( 2 \left( A + \frac{\dot{L}}{L} \right)  + Z_0 \left( Z + Y \frac{\dot{L}}{L} \right) \right) \right. \nonumber \\
&&+ \left. 2  \left( B + A \frac{\dot{L}}{L} \right) \right] \nonumber \\
N_0 &=& -2 \omega^2 + 2 \rmi \omega  \left( A + \frac{\dot{L}}{L} \right) + 2  \left( B + A \frac{\dot{L}}{L} \right).
\end{eqnarray}

This third-order differential equation fully describes the transmission coefficient $\Sto(t)$ of a superconducting lumped element microwave resonator as shown in figure \ref{fig:lemwr_schem}. To simplify this differential equation we will consider the steady state. If all parameters are constant, a steady state value $\Stoss$ of the transmission coefficient will be assumed after a sufficiently long time. In the steady state, all derivatives of $\Sto$ vanish and equation \ref{eq:S21hell02a} simplifies to:

\begin{equation} \label{eq:S21hell04}
N_1 \Stoss = N_0.
\end{equation}

To solve this expression we will make the assumption that the derivative of the inductance $L$ is small compared to $L$ itself: $\dot{L} / L \approx 0$, which then yields:

\begin{math} \label{eq:S21hell04a}
\Stoss = \frac{ -2 \frac{\omega^2}{\omegares^2} + 2 \rmi \omega \frac{L}{R} + 2 }{ -2 \frac{\omega^2}{\omegares^2} + 2 \rmi \omega \frac{L}{R} + 2       - \rmi \omega^3 Z_0 C \CCeff L - \omega^2 Z_0 \CCeff \frac{L}{R} + \rmi \omega Z_0 \CCeff }
\end{math}

By using the definitions for the macroscopic device parameters introduced previously, this result can be approximated to he steady state solution given in the main text:
\begin{equation} 
\Stoss \approx \frac{\frac{\Ql}{\Qi} + 2 \rmi \Ql \frac{f - \fres }{\fres}}{1 + 2 \rmi \Ql \frac{f - \fres }{\fres}}.
\end{equation}

Going back to the full differential equation, we can use the steady state solution in conjunction with the approximating assumption that the resonator is lossless and neglect all higher order terms of the differential equation except for the first order to get the following expression:

\begin{equation} \label{eq:S21hell05}
D \Sto = \frac{N_1}{N_2} \left( \Stoss - \Sto \right).
\end{equation}

Expressing the prefactor in the macroscopic parameters yields:

\begin{eqnarray} \label{eq:S21hell07}
\frac{N_1}{N_2}&= \rmi \omega  + \frac{2 \rmi Z_0 \omega^3 C C_C L + 2 \omega^2 \left( C + C_C \right) L + 2}{-3 Z_0 \omega^2 C C_C L + 4 \rmi \omega \left( C + C_C \right) L + Z_0 C_C}\\
&\approx \pi \dfBW + \rmi \left( \omegares - \omega \right)
\end{eqnarray}

This brings us to a first-order differential equation for the transmission parameter $\Sto$:

\begin{equation} \label{eq:S21hell09}
D\Sto(t) \approx \left( \rmi \left( \omegares(t) - \omega \right) - \pi \dfBW \right) \left( \Sto(t) - \Stoss(t) \right).
\end{equation}

Here, $\Stoss(t)$ is the steady state solutions for the parameter values at time $t$. Since the numerical methods for solving differential equations are rather demanding in terms of computation time, an iterative approximation to this is more useful for simulations. We can easily see that in the case of a constant value of $\Stoss(t) = \Stoss =const.$, the differential equation has the following solution:

\begin{equation} \label{eq:S21_firstorder}
\Sto(t) \approx \Stoss + \left( \Sto(0) - \Stoss \right) \rme^{-\pi \dfBW t +  \rmi \left( \omegares - \omega \right) t}.
\end{equation}

Even though this assumption is not true in general, it is approximately true on sufficiently small timescales. If $\Stoss(t)$ is approximately constant on a timescale $\Delta t$, we can apply expression \ref{eq:S21_firstorder} iteratively to datapoints at time intervals $\Delta t$, resulting in expression \ref{eq:S21_iter001}.
\section{Diverging readout flux noise at $\fexc = \fresO$}
\label{sec:noise_divergence}

\begin{figure}
  \includegraphics[width=0.8\columnwidth]{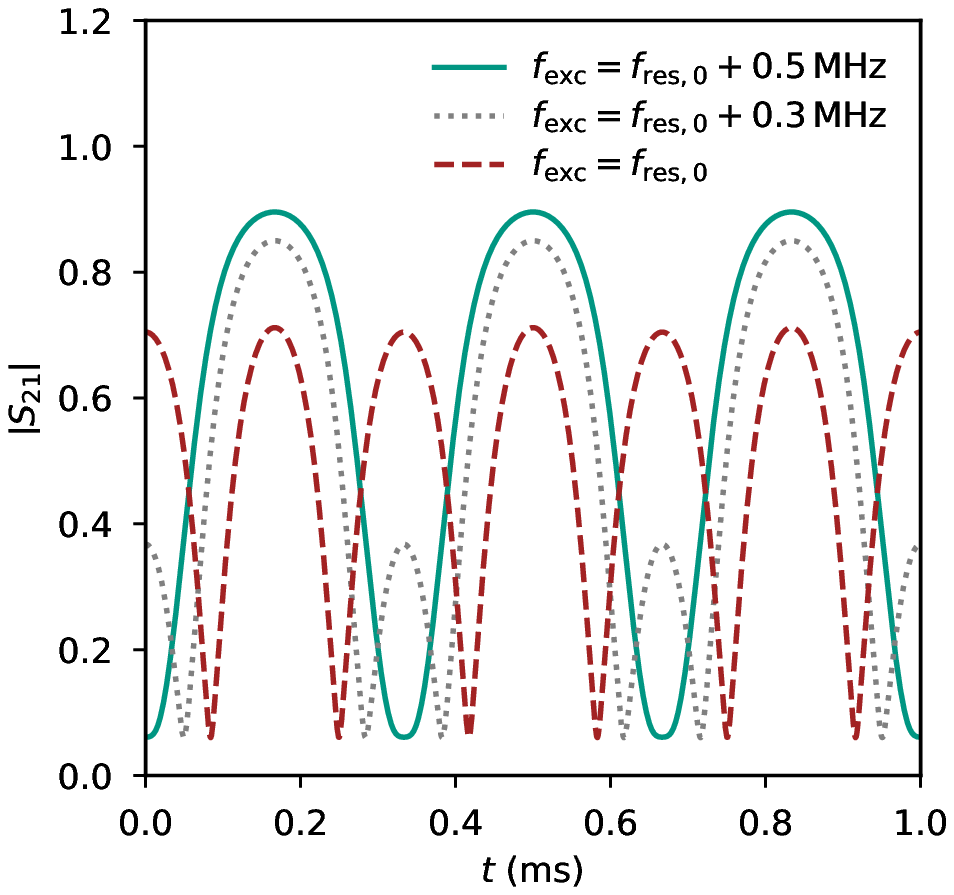}
  \caption{Transmission time traces $\left| \Sto(t) \right|$ for the special case $\betaL~<<~1$, $\Pexc \to 0$ and $\fmod << \dfBW$ for three different excitation frequencies $\fexc$. The transmission time trance for $\fexc = \fresO$ modulates with twice the frequency compared to $\fexc = \fresO + \SI{0.5}{\MHz}$.}
  \label{fig:diverging_noise}
\end{figure}

In the case of flux ramp modulation, the time-dependent resonance frequency $\fres(t)$ of a {\uMUX} channel is modulated with the modulation frequency $\fmod = \framp \Mmod \Imodmax / \PhiO$. In our simulation framework, this frequency is calculated from the device and readout parameters, and used for demodulation as shown in equation~\ref{eq:phisig_FRM}.
In the case that either $\fres(t) \leq \fexc$ or $\fres(t) \geq \fexc$ is true for all times $t$, the transmission time trace $\left| \Sto(t) \right|$ has the same periodicity as the resonance frequency $\fres(t)$. Thus, the modulation frequency $\fmod$ is the dominant frequency of the transmission time trace $\left| \Sto(t) \right|$, and the demodulation works as intended. 
However, if the excitation tone $\fexc$ is chosen very close to the unloaded resonance frequency $\fresO$, the actual resonance frequency $\fres(t)$ will change between $\fres(t) < \fexc$ and $\fres(t) > \fexc$ during flux ramp modulation. At every crossover $\fres(t) = \fexc$, a minimum in the transmission response $\left| \Sto(t) \right|$ occurs. Since this will occur twice per full period of the SQUID response, the transmission time trace $\left| \Sto(t) \right|$ now has two contributions with different modulation frequencies $\fmod$ as well as $2\fmod$ (see figure~\ref{fig:diverging_noise} grey dotted and red dashed lines). As $\fexc$ approaches $\fresO$, the contribution modulated at $\fmod$ becomes less significant. In the extreme case of $\fexc = \fresO$, $\betaL << 1$, $\Pexc \to 0$ and $\fmod << \dfBW$ as illustrated in figure~\ref{fig:diverging_noise} (red dashed line), the contribution with frequency $\fmod$ even vanishes entirely. Since the demodulation in our simulation framework is, at this point, only sensitive to the contribution with frequency $\fmod$, this leads to a significant increase in the readout flux noise around $\fexc = \fresO$.
In an experiment, one would usually demodulate the signal using the most dominant frequency present in the transmission time trace $\left| \Sto(t) \right|$. Around $\fexc = \fresO$, this would improve the readout flux noise compared to the simulation results presented in figure~\ref{fig:sim_fro} (a). In the future, a more dynamic choice of the demodulation frequency may be implemented to emulate this approach. However, as is visible in figure~\ref{fig:diverging_noise}, the transmission time trace $\left| \Sto(t) \right|$ around $\fexc = \fresO$ remains less sinusoidal and has a lower amplitude when compared to the response at a sufficiently large excitation frequency. A more sophisticated method to choose the demodulation frequency would thus not yield a readout noise lower than the global minima of figure~\ref{fig:sim_fro}.

\bibliography{99_bibliography}

\end{document}